\documentclass[11pt]{article}
\pdfoutput=1

\usepackage{bm,wrapfig,float,array,mathtools}
\usepackage{amsfonts}
\usepackage{amssymb}
\usepackage{cite}
\usepackage{amsmath}
\usepackage{color}
\usepackage{graphicx}
\usepackage{hyperref}
\usepackage{float}
\usepackage[T1]{fontenc}
\usepackage[utf8]{inputenc}
\usepackage{alphabeta}
\usepackage{fancyvrb}
\usepackage[dvipsnames]{xcolor}
\usepackage{bold-extra}
\usepackage{lmodern}
\usepackage{cleveref}
\usepackage[normalem]{ulem}

\graphicspath{ {figures/} }

\usepackage{tikz}
\usetikzlibrary{arrows}
\usetikzlibrary{shapes.geometric,calc,arrows, positioning,shapes.misc,decorations.markings}
\tikzset{
  big arrow/.style={
    decoration={markings,mark=at position 1 with {\arrow[scale=2,#1]{>}}},
    postaction={decorate},
    shorten >=0.4pt},
  big arrow/.default=black}
  
\pgfdeclarelayer{edgelayer}
\pgfdeclarelayer{nodelayer}
\pgfsetlayers{edgelayer,nodelayer,main} 
\tikzstyle{none}=[inner sep=0pt] 

\tikzstyle{NodeCross}=[draw, shape=circle, cross out, inner sep=0pt, minimum size=6pt,line width=0.25mm]
\tikzstyle{Circle}=[draw, shape=circle, black, inner sep=0pt, minimum size=6pt]
\tikzstyle{rtriangle}=[fill=black, regular polygon, regular polygon sides=3, rotate=90, inner sep=0pt, minimum size=8pt]
\tikzstyle{ltriangle}=[fill=black, regular polygon, regular polygon sides=3, rotate=270, inner sep=0pt, minimum size=8pt]
\tikzstyle{rtriangleblue}=[fill={rgb,255: red,17; green,160; blue,255}, regular polygon, regular polygon sides=3, rotate=90, inner sep=0pt, minimum size=8pt]
\tikzstyle{ltriangleblue}=[fill={rgb,255: red,17; green,160; blue,255}, regular polygon, regular polygon sides=3, rotate=270, inner sep=0pt, minimum size=8pt]
\tikzstyle{rtrianglegreen}=[fill={rgb,255: red,69; green,255; blue,28}, regular polygon, regular polygon sides=3, rotate=90, inner sep=0pt, minimum size=8pt]
\tikzstyle{ltrianglegreen}=[fill={rgb,255: red,69; green,255; blue,28}, regular polygon, regular polygon sides=3, rotate=270, inner sep=0pt, minimum size=8pt]
\tikzstyle{Uprtriangle}=[fill=black, regular polygon, regular polygon sides=3, rotate=0, inner sep=0pt, minimum size=8pt]
\tikzstyle{Downltriangle}=[fill=black, regular polygon, regular polygon sides=3, rotate=180, inner sep=0pt, minimum size=8pt]
\tikzstyle{rtriangleAmber}=[fill={rgb,255: red, 191; green, 144; blue, 63}, regular polygon, regular polygon sides=3, rotate=90, inner sep=0pt, minimum size=8pt]
\tikzstyle{UprtriangleViolett}=[fill={rgb,255: red,139; green,41; blue,148}, regular polygon, regular polygon sides=3, rotate=0, inner sep=0pt, minimum size=8pt]

\tikzstyle{Downltriangle}=[fill=black, regular polygon, regular polygon sides=3, rotate=180, inner sep=0pt, minimum size=8pt]
\tikzstyle{UpRighttriangle}=[fill=black, regular polygon, regular polygon sides=3, rotate=45, inner sep=0pt, minimum size=8pt]
\tikzstyle{UpLefttriangle}=[fill=black, regular polygon, regular polygon sides=3, rotate=315, inner sep=0pt, minimum size=8pt]
\tikzstyle{DownRighttriangle}=[fill=black, regular polygon, regular polygon sides=3, rotate=135, inner sep=0pt, minimum size=8pt]
\tikzstyle{DownLighttriangle}=[fill=black, regular polygon, regular polygon sides=3, rotate=225, inner sep=0pt, minimum size=8pt]

\tikzstyle{DashedLine}=[-, densely dashed, line width=0.25mm]
\tikzstyle{DashedLineBrown}=[-, densely dashed, line width=0.25mm, draw={rgb,255: red,155; green,103; blue,51}]
\tikzstyle{DashedLineFall}=[-, densely dashed, line width=0.25mm, draw={rgb,255: red,195; green,0; blue,0}]
\tikzstyle{DashedLineViolett}=[-, densely dashed, line width=0.25mm, draw={rgb,255: red,139; green,41; blue,148}]
\tikzstyle{DottedLine}=[-, dotted, line width=0.25mm]
\tikzstyle{BlueLine}=[-, fill=none, draw={rgb,255: red,17; green,160; blue,255}, line width=0.25mm]
\tikzstyle{GreenLine}=[-, fill=none, draw={rgb,255: red,69; green,255; blue,28}, line width=0.25mm]
\tikzstyle{RedLine}=[-, draw={rgb,255: red,191; green,0; blue,0}, fill=none, line width=0.25mm]
\tikzstyle{DashedLineRed}=[-, densely dashed, fill=none, draw={rgb,255: red,191; green,0; blue,0}, line width=0.25mm]
\tikzstyle{ThickLine}=[-, line width=0.25mm]
\tikzstyle{ViolettLine}=[-, draw={rgb,255: red,132; green,60; blue,191}, fill=none, line width=0.25mm]
\tikzstyle{ViolettDashedLine}=[-, densely dashed, draw={rgb,255: red,132; green,60; blue,191}, fill=none, line width=0.25mm]
\tikzstyle{AmberLine}=[-, draw={rgb,255: red,191; green,144; blue,63}, fill=none, line width=0.25mm]
\tikzstyle{DashedRedThick}=[-, densely dashed, fill=none, draw={rgb,255: red,191; green,0; blue,0}, line width=0.40mm]
\tikzstyle{DashedBlueThick}=[-, densely dashed, fill=none, black, line width=0.40mm]

\newcommand{\bea}{\begin{eqnarray}}
\newcommand{\eea}{\end{eqnarray}}
\newcommand{\be}{\begin{equation}}
\newcommand{\ee}{\end{equation}}
\newcommand{\ba}{\begin{aligned}}
\newcommand{\ea}{\end{aligned}}
\newcommand{\bit}{\begin{itemize}}
\newcommand{\eit}{\end{itemize}}
\newcommand{\ben}{\begin{enumerate}}
\newcommand{\een}{\end{enumerate}}
\newcommand{\nn}{\nonumber}
\renewcommand{\ni}{\noindent}

\newcommand{\tn}[1]{\textnormal{#1}}

\newcommand{\wh}{\widehat}

\newcommand{\half}{\frac{1}{2}}

\newcommand{\Z}{{\mathbb Z}}
\newcommand{\R}{{\mathbb R}}
\newcommand{\C}{{\mathbb C}}

\renewcommand{\P}{{\mathbb P}}

\newcommand{\cG}{\mathcal{G}}

\newcommand{\cL}{\mathcal{L}}

\newcommand{\cN}{\mathcal{N}}
\newcommand{\cO}{\mathcal{O}}
\newcommand{\cP}{\mathcal{P}}

\newcommand{\cS}{\mathcal{S}}

\newcommand{\F}{\mathsf{F}}
\renewcommand{\S}{\mathsf{S}}

\renewcommand{\C}{\mathsf{C}}

\renewcommand{\C}{\mathsf{C}}

\renewcommand{\L}{\mathsf{\Lambda}}

\newcommand{\fT}{\mathfrak{T}}

\newcommand{\fe}{\mathfrak{e}}
\newcommand{\ff}{\mathfrak{f}}
\newcommand{\fg}{\mathfrak{g}}

\newcommand{\su}{\mathfrak{su}}
\renewcommand{\sp}{\mathfrak{sp}}
\newcommand{\so}{\mathfrak{so}}

\newcommand{\ubf}[1]{\underline{\bf #1}}

\begin{document}

\baselineskip=18pt  
\numberwithin{equation}{section}  
\allowdisplaybreaks  






\thispagestyle{empty}

\vspace*{0.8cm} 
\begin{center}
{{\huge 1-form Symmetries of $4d$ $\mathcal{N}=2$ Class S Theories}}

 \vspace*{1.5cm}
Lakshya Bhardwaj, Max H\"ubner, Sakura Sch\"afer-Nameki\\

 \vspace*{.5cm} 
{\it  Mathematical Institute, University of Oxford, \\
Andrew-Wiles Building,  Woodstock Road, Oxford, OX2 6GG, UK}

\vspace*{0.8cm}
\end{center}
\vspace*{.5cm}

\noindent
We determine the 1-form symmetry group for any $4d$ $\cN=2$ class S theory constructed by compactifying a $6d$ $\cN=(2,0)$ SCFT on a Riemann surface with arbitrary regular untwisted and twisted punctures. The $6d$ theory has a group of mutually non-local dimension-2 surface operators, modulo screening. Compactifying these surface operators leads to a group of mutually non-local line operators in $4d$, modulo screening and flavor charges. Complete specification of a $4d$ theory arising from such a compactification requires a choice of a maximal subgroup of mutually local line operators, and the 1-form symmetry group of the chosen $4d$ theory is identified as the Pontryagin dual of this maximal subgroup. We also comment on how to generalize our results to compactifications involving irregular punctures. Finally, to complement the analysis from 6d, we derive the 1-form symmetry from a Type IIB realization of class S theories. 

\newpage

\tableofcontents


\section{Introduction}

A massive vacuum of a $4d$ theory $\fT$ is called confining if it preserves a non-trivial subgroup of the 1-form symmetry group of $\fT$ \cite{Gaiotto:2014kfa}. 
Motivated by  confinement in $4d$ $\cN=1$ theories obtained by deforming $4d$ $\cN=2$ theories that we will study in \cite{up}, 
we develop in this paper, as a precursor, the tools to determine  the 1-form symmetry of $4d$ $\cN=2$ theories.
More specifically, we consider $4d$ $\cN=2$ theories of Class S that can be obtained by compactifying $6d$ $\cN=(2,0)$ SCFTs on a Riemann surface \cite{Gaiotto:2009we}. We allow the Riemann surface to contain closed twist lines and arbitrary regular punctures which can be either untwisted or twisted. 

It is well-known that $6d$ $\cN=(2,0)$ SCFTs are classified by a Lie algebra $\fg$ of ADE-type, and that they are \emph{relative} QFTs \cite{Freed:2012bs,Witten:1998wy,Witten:2009at}, which for the purposes of this paper can be understood as follows. The $(2,0)$ theory contains dimension-2 surface operators which are not mutually local, i.e. there is an ambiguity in defining a correlation function containing two such surface operators \cite{Seiberg:2011dr}. If there is no such ambiguity, then we call the theory an \emph{absolute} QFT instead. Fusion (OPE) of these surface operators lends the set of surface operators the structure of an abelian group. Moreover, the surface operators can be screened by dynamical strings in the theory. We denote the group of surface operators modulo screening by $\wh{Z}$.

Upon compactification to $4d$, one can wrap these surface operators along various 1-cycles on the Riemann surface to generate an abelian group $\cL$ of line operators modulo screening in $4d$. The non-locality of $6d$ surface operators descends to non-locality of these $4d$ line operators. In other words, we obtain a \emph{relative} $4d$ theory upon such a compactification. To obtain an \emph{absolute} $4d$ theory $\fT$, one needs to choose a maximal subgroup $\Lambda_\fT\subset\cL$ of mutually local $4d$ line operators\footnote{The choice of $\Lambda_\fT$ is only part of the full set of choices one needs to make in order to define an absolute $4d$ $\cN=2$ theory of Class S. For example, one can obtain a group $\cL_0$  of dimension-0 and a group $\cL_2$ of dimension-2 operators in the $4d$ theory by compactifying the $6d$ surface operators along the whole Riemann surface and along a point on the Riemann surface respectively. Then the non-locality of the $6d$ surface operators descends to a non-locality between elements of $\cL_0$ and $\cL_2$, and to choose an absolute $4d$ $\cN=2$ theory, one also needs to choose subgroups $\Lambda_0$ and $\Lambda_2$ of $\cL_0$ and $\cL_2$, such that there is no non-locality between elements of $\Lambda_0$ and $\Lambda_2$. See \cite{Gukov:2020btk} for a recent discussion.}. The group $\Lambda_\fT$ can be identified with the set of charges for the 1-form symmetry group of $\fT$. In other words, the 1-form symmetry group of $\fT$ is identified as the Pontryagin dual $\wh{\Lambda}_\fT$ of $\Lambda_\fT$  \cite{Gaiotto:2014kfa}. 

Special cases of the problem explored in this paper have been discussed previously in the literature. For example, in the case where the Riemann surface $C_g$ has no punctures and no closed twist lines, the group $\cL$ was already determined in \cite{Tachikawa:2013hya} (see also the recent paper \cite{Gukov:2020btk}) to be $H_1(C_g,\wh{Z})$. For the case of $\fg=A_1$ and arbitrary $C_g$ with arbitrary number of regular punctures, this problem was discussed in \cite{Drukker:2009tz,Gaiotto:2010be}. Another situation where this problem has been discussed arises whenever there exists a degeneration limit of $C_g$ in which the $4d$ theory can be identified as a weakly coupled $4d$ gauge theory. In such a situation, one finds a canonical splitting $\cL\simeq\cL_e\times\cL_m$, where $\cL_e$ is associated to Wilson line operators and $\cL_m$ is associated to 't Hooft line operators. In such a situation, the constraint of  mutual locality can also be understood as the constraint of Dirac quantization, and choosing a way of satisfying Dirac quantization condition (i.e. a choice of $\Lambda_\fT\subset\cL_e\times\cL_m$) can be interpreted as choosing a global form of the gauge group and possible discrete theta parameters \cite{Aharony:2013hda} (see also \cite{Gaiotto:2014kfa}).
More recently work related to the higher form symmetry of 4d SCFTs and holography was studied in \cite{Bah:2020uev}. 
In the context of non-Lagrangian 4d $\mathcal{N}=2$ SCFTs of Argyres-Douglas type the 1-form symmetries were computed using the Type IIB realization using canonical singularities in \cite{Closset:2020scj, DelZotto:2020esg}, using the general observations in \cite{Morrison:2020ool, Garcia-Etxebarria:2019cnb, Albertini:2020mdx}, which are applicable more generally to geometric engineering of SCFTs in string theory. 
Many recent papers have tackled the problem of determining higher-form symmetries in lower-dimensional QFTs starting from supersymmetric QFTs in six dimensions \cite{Eckhard:2019jgg,Morrison:2020ool,Bhardwaj:2020phs,Gukov:2020btk} (see also \cite{Apruzzi:2020zot} for a related discussion), and we expect many more interesting developments in this direction. 

Our key proposal that lets us generalize the result of \cite{Tachikawa:2013hya} is that a $6d$ surface operator wrapping a cycle surrounding a regular puncture does not contribute to the set $\cL$ of $4d$ line operators (modulo screening and flavor charges). In the case of untwisted regular punctures, any $6d$ surface operator can be wrapped around the puncture, and hence according to the above proposal, untwisted regular punctures are invisible to the determination of $\cL$ and 1-form symmetry $\wh{\Lambda}_\fT$ of a $4d$ $\cN=2$ theory $\fT$ obtained after choosing a polarization $\Lambda_\fT\subset\cL$. On the other hand, twisted regular punctures do have a non-trivial influence on the calculation of $\cL$. This is because such a puncture lives at the end of a twist line which acts non-trivially on the $6d$ surface operators, and hence only the $6d$ surface operators left invariant by this action can be inserted along a loop surrounding the twisted regular puncture. Thus, according to the above proposal, a twisted regular puncture is only invisible to the $6d$ surface operators invariant under the action of the corresponding twist line. As we discuss in various examples throughout the paper, a justification for the above proposal is that $\cL$ obtained using it matches the $\cL$ obtained using the gauge theory analysis of \cite{Aharony:2013hda} (see also \cite{Gaiotto:2014kfa}) whenever there exists a limit of the compactification in which a weakly coupled $4d$ $\cN=2$ gauge theory arises 
\cite{Gaiotto:2009we,Tachikawa:2009rb,Tachikawa:2010vg,Chacaltana:2012ch,Chacaltana:2013oka,Chacaltana:2016shw,Chacaltana:2010ks,Chacaltana:2013dsj,Chacaltana:2012zy,Chacaltana:2014jba,Chacaltana:2015dat,Chacaltana:2014nya,Chacaltana:2015bna,Chacaltana:2017boe,Chcaltana:2018zag}.

Let us now discuss a subtlety that arises due to the fact that one needs to take the area of $C_g$ to zero in passing from the $6d$ theory to the $4d$ $\cN=2$ theory. One might worry that the set $\cL$ discussed might not be the true set of line operators modulo screening in the $4d$ theory. However, this worry is alleviated by the fact that in order to define the $4d$ $\cN=2$ theory one often needs to perform a non-trivial topological twist\footnote{Note that when we refer to untwisted/twisted in this paper, we usually refer to the absence/presence of outer-automorphism twist lines, not to the topologial twist.} on $C_g$, due to which one expects protected quantities to be independent of the area of $C_g$. The set $\cL$ is such a protected quantity as each element in the set can be represented by a BPS line operator in the $4d$ $\cN=2$ theory, and the screenings can also be understood in terms of BPS particles. On the other hand, in situations where one does not need to perform a non-trivial topological twist, one expects that in general $\cL$ should only be a subset of line operators (modulo screening) in the $4d$ $\cN=2$ theory. An example where $\cL$ does not capture the correct set of line operators is discussed towards the end of section \ref{NE}.

{\renewcommand{\arraystretch}{1.6}

\begin{table}
$$
\begin{array}{|c|c|c|}\hline
\text{Class S Data} & \text{Contribution to $\sqrt{\cL}$} \cr \hline \hline 
\text{untwisted handle} &  \wh{Z}(\cG) \cr \hline 
o\text{-twisted handle} &  \text{Inv}(\wh{Z}(\cG), o) \cr \hline 
\text{untwisted regular puncture} &  0 \cr \hline 
\text{(open $\Z_2$ twist line of type $o$, open $\Z_2$ twist line of type $o$)} &  \wh{Z}(\cG)/\text{Inv}(\wh{Z}(\cG), o)  \cr \hline \hline
\text{(open $b$ line, $a$-twisted handle)} &  \Z_2 \cr \hline 
\text{(open $b$ line, open $b'$ line)} &  0 \cr \hline 
\text{(meson, meson)} &  \Z_2\times\Z_2 \cr \hline 
\text{(meson, $b$-twisted handle)} &  \Z_2 \cr \hline 
\text{(open $b$ line, meson)} &  \Z_2 \cr \hline 
\text{(baryon, baryon)} &  \Z_2\times\Z_2 \cr \hline 
\text{(meson, baryon)} &  \Z_2\times\Z_2 \cr \hline 
\text{(baryon, $b$-twisted handle)} &  \Z_2 \cr \hline 
\text{(open $b$ line, baryon)} &  \Z_2 \cr \hline 
\text{(open $b$ line, mixed configuration)} &  \Z_2 \cr \hline 
\end{array}
$$
\caption{Summary of class S data and their impact on the defect group $\cL$. The contribution is always squared, so we only list half of the contribution to $\cL$ for each kind of Class S datum. For example, the first entry describes that an untwisted handle of the Riemann surface contributes $\wh{Z}(\cG)\times\wh{Z}(\cG)$ to $\cL$. 
The first four entries are universal for any class S construction -- including contributions from the genus, punctures and twist-lines. 
Here $\wh{Z}(\cG)$ is the Pontryagin dual of the center $Z(\cG)$ of the simply connected group $\cG$ associated to the ADE-algebra $\fg$ of the $6d$ $(2,0)$ theory. 
$\text{Inv}(\wh{Z}(\cG), o)$ is the subgroup of $\wh{Z}(\cG)$ left invariant by the action of outer-automorphism $o$ on $\wh{Z}(\cG)$. An $o$-twisted handle refers to a handle carrying a closed $o$-twist line wrapped along either $A$ or $B$ cycle of the handle. An untwisted puncture does not contribute anything to $\cL$. 
The entries after the double-line refer to the $S_3$ twisted compactifications of $D_4$ $(2,0)$ theory, where open twist lines form a variety of irreducible configurations (meson, baryon etc.) and this comprises a summary of our findings in section \ref{S3open}, and we refer the reader there for a detailed discussion. 
\label{RD}}
\end{table}}

Many class S theories have known realizations in terms of local Calabi-Yau compactifications in Type IIB\footnote{{Although in principle any class S theory should have a IIB compactification associated to it, the precise construction in particular in the case of non-diagonalizable Higgs fields and irregular punctures is -- to our knoweldge -- not developed.}} in terms of an ALE-fibration over the curve $C_{g,n}$. The defect group in those cases are computed from the relative homology three-cycles of the non-compact Calabi-Yau, or equivalently, the second homology of the link (i.e. the boundary five-fold). From the local Higgs bundle realization of the ALE-fibration of the Calabi-Yau three-fold, we determine these homology groups and confirm the defect group for the case of no punctures and for regular untwisted and twisted punctures: 
The defect group $\cL$ has a simple description, purely in terms of the data on the boundary of the non-compact Calabi-Yau threefold, namely the boundary $B_F= S^3/\Gamma_{\text{ADE}} \rightarrow C$ fibration, where $C$ is the Gaiotto curve, and the base of the ALE-fibration. Then the defect group is simply given in terms of the 2-cycles of $B_F$, which extend trivially to the Calabi-Yau. 

In fact, as we discuss in section \ref{JKP}, this approach can be viewed to provide a justification for our key proposal that a $6d$ surface operator wrapping a cycle surrounding a regular puncture does not contribute to $\cL$. 
Moreover, this approach might shed light on the irregular punctures as, e.g. generalized AD theories have a realization in terms of Type IIB on canonical singularities, from which in turn the 1-form symmetry can be computed \cite{Closset:2020scj, DelZotto:2020esg}.

We find that $\cL$ can be roughly constructed from the various kinds of data on the Riemann surface used for compactification. We collect this rough decomposition of $\cL$ in Table \ref{RD} to be used as a reference. It is important to note that the table only captures the group-structure of $\cL$, while one of the key ingredients is the pairing on $\cL$ capturing the mutual non-locality of $4d$ line operators. This pairing is required to choose a polarization $\Lambda$ and determine the corresponding 1-form symmetry $\wh{\Lambda}$. The explicit form of the pairing can be found in the main text.

The paper is organized as follows. In section \ref{6d} we review some properties of dimension-2 surface operators and outer-automorphism discrete 0-form symmetries in $6d$ $\cN=(2,0)$ SCFTs. In section \ref{w/op} we discuss 1-form symmetry in absolute $4d$ $\cN=2$ theories obtained by compactifying $6d$ $(2,0)$ theories on a genus $g$ Riemann surface in the presence of arbitrary twists by outer-automorphism discrete 0-form symmetries, but without involving any punctures. In section \ref{w/p} we extend our analysis of previous section to includ arbitrary untwisted and twisted regular punctures. In section \ref{irr} we sketch how our analysis can be extended to include irregular punctures, giving explicit results for a specific class of irregular punctures of $A_{n-1}$ $(2,0)$ theories. Finally, in section \ref{sec:IIB} we argue from a Type IIB realization of class S theories for the 1-form symmetries. Our notation is summarized in appendix \ref{app:Not}.

\section{Surface Operators and Outer Automorphisms in $6d$ $(2,0)$}\label{6d}
$6d$ $\cN=(2,0)$ SCFTs are relative QFTs classified by a simple Lie algebra $\fg$ of $A,D,E$ type. Such a theory contains surface defect operators of dimension 2. Modulo screening by dynamical objects, these operators can be classified by the Pontryagin dual $\wh{Z}(\cG)$ of the center $Z(\cG)$ of the simply connected group $\cG$ associated to $\fg$, which are summarized in table \ref{tab:Pontry}. The Pontryagin dual $\widehat Z(\cG):=\tn{Hom}(Z\big(\cG),\R/\Z\big)$ of a finite abelian center group is isomorphic to the center group itself.

{\renewcommand{\arraystretch}{1.4}

\begin{table}
$$
\begin{array}{|c|c|c|c|c|}\hline
\fg &{Z}(\cG)&\wh{Z}(\cG)   &   \langle\cdot,\cdot\rangle \cr \hline \hline 
A_{n-1} &  \Z_n & \Z_n  & \langle f,f\rangle= {1\over n} \cr \hline 
D_{4n} &  \Z_2\times\Z_2 & \Z_2\times\Z_2  &	\langle s,s\rangle=0,~~\langle c,c\rangle=0,~~\langle s,c\rangle=\half 	\cr \hline
D_{4n+1} &  \Z_4 & \Z_4  &	\langle s,s\rangle={3\over 4}	\cr \hline
D_{4n+2} & \Z_2\times \Z_2 &\Z_2\times\Z_2& \langle s,s\rangle=\half,~~\langle c,c\rangle=\half,~~\langle s,c\rangle=0   \cr \hline
D_{4n+3} & \Z_4 & \Z_4 & \langle s,s\rangle={1\over 4}	\cr \hline
E_6 &\Z_3 & \Z_3&  \langle f,f\rangle= {2\over 3}	\cr \hline 
E_7& \Z_2& \Z_2 &\langle f,f\rangle= {1\over 2}		\cr \hline 
E_8 &  0 &  0 &  -	\cr \hline 
\end{array}
$$
\caption{For the ADE Lie algebras $\mathfrak{g}$ we denote by $\mathcal{G}$ the simply-connected Lie group, and list the center $Z(\cG)$, the 
Pontryagin dual group to the center $\wh{Z}(\cG)$, and the bihomomorphism $\langle\cdot,\cdot\rangle$. $E_8$ has a trivial center group, which has been denoted by 0 since we use an additive notation for the group multiplication law throughout this paper. We denote a generator of $\wh{Z}(\cG)$ for $\fg=A_{n-1},E_6,E_7$ as $f$; a generator of $\wh{Z}(\cG)$ for $\fg=D_{2n+1}$ as $s$; and generators of $\wh{Z}(\cG)\simeq\Z_2\times\Z_2$ for $\fg=D_{2n}$ as $s,c$. We also define $v:=s+c$ for $\fg=D_{2n}$.\label{tab:Pontry}}
\end{table}}

These surface operators are not all mutually local. Consider a correlation function containing two surface operators $\alpha,\beta\in\wh{Z}(G)$. As $\alpha$ is moved around $\beta$, the correlation function is transformed by a phase factor\footnote{Notice (in the following equation) that we define the pairing with a negative sign as compared to the standard choice, which can be found for example in \cite{Gukov:2020btk}.} 
\be
\text{exp}\big(2\pi i \langle\alpha,\beta\rangle\big)
\ee
with a bihomomorphism
\be\label{eq:BiHom}
\langle\cdot,\cdot\rangle:\quad  \wh{Z}(\mathcal{G})\times \wh{Z}(\mathcal{G})\  \to\  \R/\Z \,.
\ee 
The bihomomorphism can be specified by providing its values on the generators of $\wh{Z}(G)$ \cite{Gukov:2020btk}. These are also listed in table \ref{tab:Pontry}.

The $(2,0)$ theory admits a discrete 0-form symmetry which can be identified with the group of outer-automorphisms $\cO_\fg$ of $\fg$, which are
\be
\cO_\fg = \Z_2
\ee
for $\fg=A_{n\ge2},D_{n\ge5},E_6$, and
\be
\cO_{D_4} = S_3 \,,
\ee
namely the group formed by permutations of three objects. $\cO_\fg$ is trivial for $E_7$ and $E_8$. The outer-automorphisms act on representations of $\fg$, and hence on $\wh{Z}(\cG)$. For $\fg=A_n,D_{2n+1},E_6$, the non-trivial element of $\cO_\fg=\Z_2$ acts by sending the generator of $\wh{Z}(\cG)$ to its inverse. For $\fg=D_{2n}$ and $n\ge3$, the non-trivial element of $\cO_\fg=\Z_2$ acts by exchanging the two chosen generators $s,c$ of $\wh{Z}(\cG)\simeq \Z_2\times\Z_2$. For $\fg=D_4$, we generate $\cO_{D_4} = S_3$ in terms of a $\Z_3$ and a $\Z_2$ subgroup of it. We choose generators $a\in\Z_3$ and $b\in\Z_2$, which act as follows
\be\ba
a:\qquad &s\to v,~~v\to c,~~c\to s\cr 
b: \qquad& s\to c,~~c\to s,~~v\to v  \,.
\ea
\ee
Then the elements of $S_3$ can be written as $1,a,a^2,b,ab,a^2b$. An important conjugation relation we will use throughout the paper is $bab=a^2$.

\section{Compactifications without Punctures} 
\label{w/op}

In this section we consider compactifications of 6d $(2,0)$ theories on a Riemann surface $C_g$ of genus $g$ without any punctures. If there are no other ingredients involved in the compactification, such a compactification is called as an \emph{untwisted} compactification. On the other hand, we can also consider \emph{twisted} compactifications which means the following. The outer-automorphism 0-form symmetry in $6d$ $(2,0)$ theory discussed in the last section is generated by topological operators of codimension-1 in the $6d$ theory. Inserting such a topological operators along a cycle of the Riemann surface gives rise to a ``codimension-0 object'' in the $4d$ theory, which means that the resulting $4d$ theory itself is different from the $4d$ theory arising when no such topological operators are inserted. We often refer to the locus of the topological operator on $C_g$ as a \emph{twist line}, and when this locus is a 1-cycle on $C_g$ we say that the twist line is \emph{closed}.  In the presence of punctures this picture is enhanced by the alternative of \emph{open} twist lines. Open twist lines emanate and end at punctures and we discuss their effect in section \ref{w/p}.

Twisted and untwisted compactifications can equivalently be distinguished in the Higgs bundle description of the compactification. Here the insertion of topological operators along twist lines gives rise to an action on the Higgs field by an outer automorphism $o$ across these. The insertions alter the gauge group of the effective 4d $\cN=2$ theory and have a geometric interpretation in the IIB dual description as we explain in more detail in section \ref{sec:IIB}. In this geometric picture we are further able to justify the key assumption that regular untwisted punctures are irrelevant in determining the defect group, which we also argue for in the section \ref{w/p}.

\subsection{Untwisted Case}

Let us compactify a $(2,0)$ theory on a Riemann surface $C_g$ of genus $g$ without any punctures or twists. This gives rise to a relative $4d$ $\cN=2$ theory with a set of line defects descending from the elements of $\wh{Z}(\cG)$ wrapped along various cycles of $C_g$. That is, the set $\cL$ of $4d$ line defects (modulo screening) can be identified with
\be
H_1(C_g,\wh{Z})\simeq H_1(C_g,\Z)\otimes\wh{Z}\,.
\ee
These line defects are not all mutually local. The violation of mutual locality between two elements $a\otimes\alpha,b\otimes\beta\in H_1(C_g,\Z)\otimes\wh{Z}\simeq H_1(C_g,\wh{Z})$ is captured by the phase
\be\label{p}
\text{exp}\big(2\pi i \langle\alpha,\beta\rangle\langle a,b\rangle\big)\,,
\ee
where $\langle a,b\rangle$ is the intersection pairing on $H_1(C_g,\Z)$. This gives rise to a pairing on $H_1(C_g,\wh{Z})$ which is the natural combination of the intersection pairing and the bihomomorphism \eqref{eq:BiHom}  
\be\label{eq:Pairing}
\ba
\langle\cdot,\cdot\rangle\,: \qquad & H_1(C_g,\wh{Z})\times H_1(C_g,\wh{Z})\, \rightarrow \, \R/\Z \cr 
& \langle a \otimes \alpha,b \otimes \beta \rangle= \langle a,b\rangle \langle \alpha,\beta\rangle \,.
\ea
\ee
We can specify an absolute $4d$ $\cN=2$ theory by choosing a maximal set of line operators 
\be
\Lambda\subset H_1(C_g,\wh{Z})\,,
\ee
which are all mutually local, i.e. the phase (\ref{p}) is trivial for any two elements in $\Lambda$. Such a set $\Lambda$ is also referred to as a `maximal isotropic subgroup' or as a `polarization' in what follows. The 1-form symmetry of the absolute $4d$ $\cN=2$ theory can then be identified with the Pontryagin dual $\wh{\Lambda}$ of $\Lambda$.

Once we choose a set of A and B cycles on $C_g$, we can decompose
\be\label{eq:ABdecomp}
H_1(C_g,\wh{Z})\simeq \wh{Z}_A^g\times\wh{Z}^g_B\,,
\ee
where $\wh{Z}^g_A$ is the contribution of A-cycles, and $\wh{Z}^g_B$ is the contribution of B-cycles. Moreover, $\wh{Z}^g_A$ and $\wh{Z}^g_B$ are maximal isotropic sublattices, and hence provide canonical choices of $\Lambda$ once a choice of A and B cycles has been made.

\vspace{6mm}

\ni \ubf{Example}: When $(2,0)$ theory of type $\fg$ is compactified on a torus, we obtain $4d$ $\cN=4$ SYM with gauge algebra $\fg$. Choosing an A-cycle and a B-cycle, we write
\be
H_1(T^2,\wh{Z})\simeq \wh{Z}_A\times\wh{Z}_B\,.
\ee
We assume without loss of generality that the A-cycle is much shorter than the B-cycle. Then, $\wh{Z}_A$ can be identified as the set of $4d$ Wilson line operators, and $\wh{Z}_B$ can be identified as the set of $4d$ 't Hooft line operators. Choosing $\Lambda=\wh{Z}_A$, we obtain $4d$ $\cN=4$ SYM with gauge group $\cG$. On the other hand, choosing $\Lambda=\wh{Z}_B$, we obtain $4d$ $\cN=4$ SYM with gauge group $\cG/Z(\cG)$ and all discrete theta parameters turned off. In these cases, we have 1-form symmetry
\be
\wh{\Lambda}\simeq Z(\cG)
\,, \ee
which matches with the 1-form symmetry obtained using the Lagrangian description of $4d$ $\cN=4$ SYM: when the gauge group is $\cG$, this is the electric 1-form symmetry; and then the gauge group is $\cG/Z(\cG)$, this is the magnetic 1-form symmetry.

Other choices of global forms of the gauge group and discrete theta angles are obtained by choosing other polarizations. For concreteness, consider the case of $\fg=\su(4)$. In this case, $\wh{Z}_A\simeq\wh{Z}_B\simeq\Z_4$. The $PSU(4)$ theory with a discrete theta parameter $n\in\{0,1,2,3\}$ turned on is obtained by choosing $\Lambda$ to be the sublattice generated by the element $(n,1)\in\Z_4\times\Z_4\simeq \wh{Z}_A\times\wh{Z}_B$ (where we have represented $\Z_4$ as the additive group $\Z/4\Z$). Any such choice leads to the 1-form symmetry
\be
\wh{\Lambda}\simeq \Z_4
\,.
\ee
If we choose the polarization $\Lambda$ generated by elements $(0,2)$ and $(2,0)$ in $\Z_4\times\Z_4$, then we obtain the $SO(6)\simeq SU(4)/\Z_2$ theory with the discrete theta parameter turned off. In this case the 1-form symmetry is
\be
\wh{\Lambda}\simeq \Z_2\times\Z_2
\,. \ee
From the point of view of the Lagrangian description, the two $\Z_2$ factors are electric and magnetic 1-form symmetries respectively. The remaining $\su(4)$ theory has $SO(6)$ gauge group and a discrete theta parameter turned on. This is obtained by choosing $\Lambda$ to be generated by the element $(1,2)\in\Z_4\times\Z_4\simeq \wh{Z}_A\times\wh{Z}_B$, and the 1-form symmetry group of the theory is
\be
\wh{\Lambda}\simeq \Z_4 \,.
\ee

\vspace{6mm}

\ni \ubf{Example}: Consider compactifying $A_1$ $(2,0)$ theory on $C_g$ with $g\ge2$. In an S-duality frame, in which A-cycles are much shorter than B-cycles, we obtain the following Lagrangian $4d$ $\cN=2$ theory
\be
\begin{tikzpicture}
\node (v1) at (-0.45,0.9) {$\so(4)$};
\begin{scope}[shift={(-3.6,0)}]
\node (v-1) at (-0.45,0.9) {$\so(3)$};
\end{scope}
\begin{scope}[shift={(-1.8,0)}]
\node (v0) at (-0.4,0.9) {$\su(2)$};
\end{scope}
\begin{scope}[shift={(1.8,0)}]
\node (v2) at (-0.45,0.9) {$\su(2)$};
\end{scope}
\begin{scope}[shift={(3.6,0)}]
\node (v3) at (-0.45,0.9) {$\so(4)$};
\end{scope}
\begin{scope}[shift={(4.9,0)}]
\node (v4) at (-0.45,0.9) {$\cdots$};
\end{scope}
\begin{scope}[shift={(6.2,0)}]
\node (v5) at (-0.45,0.9) {$\so(4)$};
\end{scope}
\begin{scope}[shift={(8,0)}]
\node (v6) at (-0.4,0.9) {$\su(2)$};
\end{scope}
\begin{scope}[shift={(9.8,0)}]
\node (v7) at (-0.45,0.9) {$\so(3)$};
\end{scope}
\draw  (v0) edge (v1);
\draw  (v1) edge (v2);
\draw  (v2) edge (v3);
\draw  (v3) edge (v4);
\draw  (v4) edge (v5);
\draw  (v5) edge (v6);
\draw  (v6) edge (v7);
\draw  (v-1) edge (v0);
\node (v8) at (-2.2,2.1) {$\half\F$};
\draw  (v8) edge (v0);
\node (v9) at (7.6,2.1) {$\half\F$};
\draw  (v9) edge (v6);
\end{tikzpicture}
\ee
where we have a total of $2g-1$ nodes. Each node describes a gauge algebra and an edge between two nodes denotes a half-bifundamental\footnote{Here, for ease of notation, we are using the convention that the fundamental representation of $\so(n)$ is the $n$-dimensional vector representation. So, the fundamental representation for $\so(3)$ is not the fundamental representation for $\su(2)$, but rather the adjoint representation. Similarly, the fundamental representation of $\so(4)$ is the $(2,2)$ rep of $\su(2)\oplus\su(2)\simeq\so(4)$.} between the two nodes. An edge connecting an $\su(2)$ node to a node labeled $\half\F$ implies that the corresponding $\su(2)$ gauge algebra carries an extra half-hyper charged in fundamental rep. If we choose $\Lambda=(\Z/2\Z)_A^g$, we obtain the $4d$ theory with all the gauge groups being simply connected. In this case, we have 1-form symmetry
\be
\wh{\Lambda}\simeq \Z_2^g\,,
\ee
which can be easily matched with the above Lagrangian description with all the gauge groups chosen to be the simply connected ones. A $\Z_2$ factor arises from each of the $g$ number of $\so(n)$ nodes (where $n=3,4$ and the corresponding gauge group is Spin$(n)$). This $\Z_2$ is the subgroup of center of Spin$(n)$ that acts trivially on the fundamental representation of $\so(n)$ as defined in the above footnote.

\subsection{Including Closed $\Z_2$ Twist-lines}\label{NE}
We can also consider twisted compactifications of $6d$ $\cN=(2,0)$ on $C_g$ (without punctures). This involves wrapping the topological defects generating the outer-automorphism discrete 0-form symmetries along cycles on $C_g$. In this subsection we either consider those $\fg$ for which the outer-automorphism group is $\Z_2$, or the case $\fg=D_4$ with twist lines valued only in the $\Z_2$ subgroup of the $S_3$ outer-automorphism group generated by the element $b$ (see section \ref{6d}). We can wrap the $\Z_2$ twist lines along some $L\in H_1(C_g,\Z_2)$. Let us first discuss the case of $g=1$. Without loss of generality we can choose $L$ to be the B-cycle of the torus.

\begin{figure}
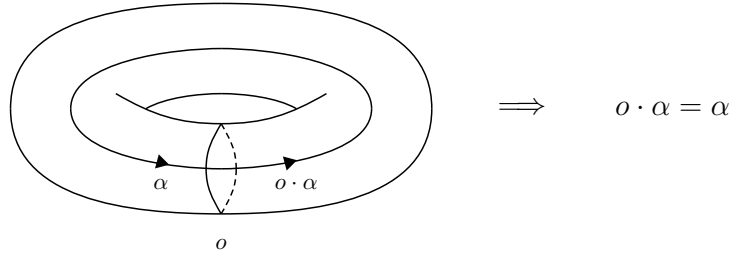

\centering
\scalebox{.8}{
}
\caption{A closed $\Z_2$ twist line $o$ is inserted along the B-cycle of a torus. An element $\alpha\in\wh{Z}$ inserted along the A-cycle is acted upon by $o$ as it crosses the closed twist line. Since the A-cycle closes back to itself we deduce that only the elements $\alpha$ left invariant by the action of $o$ can be inserted along the A-cycle.}
\label{fig:InvariantALine}
\end{figure}

Then, along the dual A-cycle, we can only wrap those elements of $\wh{Z}$ which are left invariant by the action of $\Z_2$ outer-automorphism $o$ -- see figure \ref{fig:InvariantALine}. 
Let us denote this subgroup of $\wh{Z}$ by 
\be\label{eq:Inv}
\text{Inv}(\wh{Z}, o):= \wh{Z}|_o = \{z \in \wh{Z}: \ o \cdot z = z \} \,.
\ee
For $\fg=A_{2n-1}$, we can only wrap the element $nf\in\wh{Z}\simeq\Z_{2n}$ and hence 
\be
\text{Inv}(\wh{Z}, o)\simeq\Z_2 \,.
\ee
Similarly, for $\fg=D_{n}$, only $v$ can be wrapped and hence 
\be
 \text{Inv}(\wh{Z}, o)\simeq\Z_2\,.
\ee
For $\fg=A_{2n}$ and $\fg=E_6$, no element in $\wh{Z}$ can be wrapped and hence $ \text{Inv}(\wh{Z}, o)$ is trivial. For $\fg=E_7,E_8$ the group of outer automorphisms is trivial.

\begin{figure}
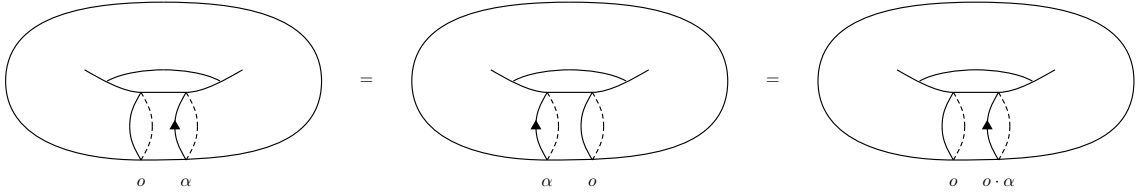

\centering
\scalebox{.6}{

}
\caption{A closed $\Z_2$ twist line $o$ is inserted along the B-cycle of a torus. An element $\alpha\in\wh{Z}$ inserted along the B-cycle can be moved around and converted to the element $o\cdot \alpha$ inserted along the B-cycle.}
\label{fig:CommutatorOAlpha}
\end{figure}

On the other hand, along the B-cycle we can wrap any element $\alpha\in\wh{Z}$, but moving it across the twist line implies that $\alpha$ can be identified with the element $o\cdot\alpha\in\wh{Z}$ where $o\cdot\alpha$ is obtained by applying the $\Z_2$ action on $\alpha$. See figure \ref{fig:CommutatorOAlpha}. The set of $4d$ line operators descending from the B-cycle, which we denote as
\be\label{eq:EquivClass}
{{\text{Proj}(\wh{Z},o)} :={\wh{Z} \over \left\langle  g- o \cdot g\right\rangle }}
\ee
 can be obtained by modding out $\wh{Z}$ by the identifications imposed by $o$, that is by modding out $\wh{Z}$ by the subgroup $\langle g- o \cdot g\rangle\subseteq\wh{Z}$ generated by the element $g- o \cdot g\in\wh{Z}$ where $g$ is a generator of $\wh{Z}$. For $\fg=D_{n}$, the action of $o$ implies that $s\sim c$, which implies $v\sim 0$, and consequently 
\be
\text{Proj}(\wh{Z},o)\simeq\Z_2 \,,
\ee
whose non-trivial element can be identified either with $s$ or with $c$. For $\fg=A_{n-1}$ and $\fg=E_6$, we have $f\sim -f$, which implies that $2mf\sim0$ for all $m\in\Z$. Thus for $\fg=A_{2n-1}$, we have 
\be
\text{Proj}(\wh{Z},o)\simeq\Z_2 \,,
\ee
whose non-trivial element can be represented by any element of the form $(2m+1)f\in\wh{Z}\simeq\Z_{2n}$. For $\fg=A_{2n}$ and $\fg=E_6$, we can write $f=2mf$ for $m=n+1$ and $m=2$ respectively, and hence $\text{Proj}(\wh{Z},o)$ is trivial.

Now, notice that $\text{Inv}(\wh{Z}, o)$ and $\text{Proj}(\wh{Z},o)$ have a non-trivial mutual pairing which descends from the mutual pairing \eqref{eq:Pairing} between $\wh{Z}_{A}$ and $\wh{Z}_{B}$. For example, for $\fg=D_n$, the generator for $\text{Inv}(\wh{Z}, o)\simeq\Z_2$ is $v\in\wh{Z}_{A}$, and the generator for $\text{Proj}(\wh{Z},o)\simeq\Z_2$ can be taken to be $s\in\wh{Z}_{B}$. Then the pairing between the generators is
\be
\langle v,s\rangle=\half\,.
\ee
Had we chosen the generator of $\text{Proj}(\wh{Z},o)$ to be $c\in\wh{Z}_{B}$ instead, we would have obtained the same pairing as above. For $\fg=A_{2n-1}$, the generator for $\text{Inv}(\wh{Z}, o)\simeq\Z_2$ is $nf\in\wh{Z}_{A}$, and the generator for $\text{Proj}(\wh{Z},o)\simeq\Z_2$ can be taken to be some $(2m+1)f\in\wh{Z}_{B}$. The pairing between the generators is
\be
\langle nf,(2m+1)f\rangle=\half \,,
\ee
irrespective of the value of $m$.

An absolute $4d$ $\cN=2$ theory is then specified by choosing
\be
\Lambda\subset\cL\simeq\text{Inv}(\wh{Z}, o) \times \text{Proj}(\wh{Z},o) \,,
\ee
with $\Lambda$ being maximally isotropic. The 1-form symmetry of the $4d$ $\cN=2$ theory can then be identified with $\wh{\Lambda}$.

\begin{figure}
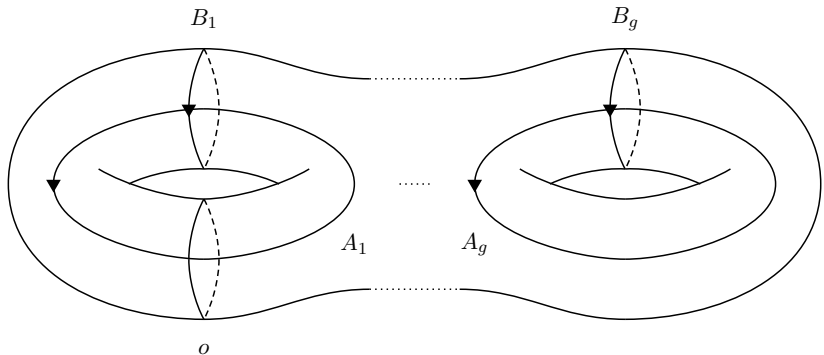

\centering
\scalebox{.8}{
}
\caption{A Riemann surface of genus $g$ with a closed $\Z_2$ twist line $o$ wrapped along the $B_1$ cycle. }
\label{fig:ClosedTwistLineWLOG}
\end{figure}

For a general $C_g$ with arbitrary $g$, the twist lines are specified by picking an element $L\in H_1(C_g,\Z_2)$. By Poincare duality, we can work with the dual element $\wh{L}\in H^1(C_g,\Z_2)$. Choose a set of A and B cycles on $C_g$. Then $\wh{L}$ assigns values $\wh{L}(A_i),\wh{L}(B_i)\in\{0,1\}$ to all cycles $A_i, B_i$. We can perform $Sp(2g,\Z)$ transformations to transform to a new set of A and B cycles such that only $\wh{L}(A_1)=1$, while $\wh{L}(A_i)=0$ for all $i\neq 1$ and  $\wh{L}(B_i)=0$ for all $i$. That is, in this frame, which can always be chosen, the twist line $L$ wraps only the cycle $B_1$. See figure \ref{fig:ClosedTwistLineWLOG}.

Now, combining the results discussed previously, we easily identify the set $\cL$ of $4d$ line operators (modulo screening). We find that the polarization is chosen as
\be\label{Z2t}
\Lambda\subset\cL\ \simeq\ \text{Inv}(\wh{Z}, o) \times \text{Proj}(\wh{Z},o)\times \wh{Z}^{g-1}_A \times \wh{Z}^{g-1}_B\,,
\ee
where the pairing is obvious from our previous discussion. The 1-form symmetry group of such an absolute $4d$ $\cN=2$ theory is identified as $\wh{\Lambda}$.

\vspace{6mm}

\ni\ubf{Example}: Consider compactifying $D_{n+1}$ $(2,0)$ theory on a torus, and wrap a $\Z_2$ twist line along the B-cycle. We can write the set of line defects as
\be\label{eq:ExampleT2}
\text{Inv}(\wh{Z}, o) \times \text{Proj}(\wh{Z},o)\simeq(\Z/2\Z)_A\times(\Z/2\Z)_B\,.
\ee
First, assume that the A-cycle is much shorter than the B-cycle. This corresponds to first compactifying $D_{n+1}$ $(2,0)$ theory on a circle with outer-automorphism twist, leading to $5d$ $\cN=2$ SYM with $\sp(n)$ gauge algebra and discrete theta angle $\theta=0$ \cite{Tachikawa:2011ch}. We further compactify this $5d$ theory on another circle obtaining $4d$ $\cN=4$ SYM with $\sp(n)$ gauge algebra. Choosing $\Lambda=(\Z/2\Z)_A$ corresponds to picking the simply connected $Sp(n)$ gauge group for the $4d$ theory, and the 1-form symmetry $\wh{\Lambda}\simeq\Z_2$ can be identified as the electric 1-form symmetry from the point of view of this Lagrangian $4d$ theory. Choosing $\Lambda=(\Z/2\Z)_B$ leads to gauge group $Sp(n)/\Z_2$ with the discrete theta parameter turned off, and the 1-form symmetry $\wh{\Lambda}\simeq\Z_2$ can be identified as the magnetic 1-form symmetry from the point of view of this Lagrangian $4d$ theory.

Now, assume that the B-cycle is much shorter than the A-cycle. This corresponds to first compactifying $D_{n+1}$ $(2,0)$ theory on a circle \emph{without} outer-automorphism twist, leading to $5d$ $\cN=2$ SYM with $\so(2n+2)$ gauge algebra. We further compactify this $5d$ theory on another circle with a $\Z_2$ outer-automorphism twist, leading to $4d$ $\cN=4$ SYM with $\so(2n+1)$ gauge algebra. Choosing $\Lambda=(\Z/2\Z)_A$ corresponds to picking the $SO(2n+1)$ gauge group for the $4d$ theory with all discrete theta parameters turned off, and the 1-form symmetry $\wh{\Lambda}\simeq\Z_2$ can be identified as the magnetic 1-form symmetry from the point of view of this Lagrangian $4d$ theory. Choosing $\Lambda=(\Z/2\Z)_B$ leads to the simply connected gauge group Spin$(2n+1)$, and the 1-form symmetry $\wh{\Lambda}\simeq\Z_2$ can be identified as the electric 1-form symmetry from the point of view of this Lagrangian $4d$ theory.

\vspace{6mm}

\ni\ubf{Non-example}: Consider compactifying $A_{2n}$ $(2,0)$ theory on a torus, and wrap a $\Z_2$ twist line along the B-cycle. Our proposal would predict the set $\cL$ of $4d$ line defects (modulo screening) to be
\be
\cL\simeq\text{Inv}(\wh{Z}, o) \times \text{Proj}(\wh{Z},o)\simeq0\,,
\ee
which is the trivial group. That is, all the $4d$ line defects are proposed to be screened. Correspondingly, the 1-form symmetry of the resulting $4d$ theory is predicted to be trivial. These predictions are incorrect as we now show.

The limit for which the A-cycle is much shorter than the B-cycle corresponds to first compactifying the $A_{2n}$ $(2,0)$ theory on a circle of radius $R_6$ with an outer-automorphism twist, thus leading to $5d$ $\cN=2$ SYM with $\sp(n)$ gauge algebra  with gauge coupling $g_{YM}^2= R_6$ and discrete theta angle $\theta=\pi$ \cite{Tachikawa:2011ch}. Due to the presence of non-trivial discrete theta angle the BPS instanton particle in this $5d$ theory transforms in the fundamental representation of $\sp(n)$. Thus, the group of line operators (modulo screening) in this $5d$ theory is trivial. Moreover, every possible 't Hooft dimension-2 surface operator in the $5d$ theory which is local with the above mentioned instanton BPS particle is screened. Thus, the group of surface operators (modulo screening) in this $5d$ theory is also trivial.

Compactifying the above $5d$ theory further on a circle of finite non-zero radius $R_5$, one expects the $4d$ theory obtained to have no line defects (modulo screening), since there are no line or surface defects (modulo screening) in the $5d$ theory as we saw above. This is so far consistent with our above predictions.

However, as we send $R_5,R_6\to0$ while keeping $R_6/R_5$ preserved, we obtain the $4d$ $\cN=4$ theory having $\fg=\sp(n)$ with gauge coupling $g_{YM}^2= R_6/R_5$ and theta angle $\theta=\pi$. This $4d$ theory clearly has a $\Z_2\times\Z_2$ group of $4d$ line operators (modulo screening). Thus, our above predictions do not provide the correct answer in the limit when the torus is shrunk to zero size.

From the point of view of the above $5d$ theory, this limit decouples the BPS instanton particle responsible for screening the fundamental Wilson line, since the mass $m$ of the BPS instanton particle scales as $m\sim 1/R_6\to\infty$. This means that the fundamental Wilson line is not screened after taking this limit. Moreover, the 't Hooft operator which was not mutually local with the BPS instanton particle becomes available, and we recover the correct result that the set of $4d$ line operators (modulo screening) is $\Z_2\times\Z_2$. There are 3 distinct choices of polarization corresponding to choosing the $4d$ gauge group $Sp(n)$ and $Sp(n)/\Z_2$ with a discrete $\Z_2$ valued theta parameter. In each of the three cases, the true 1-form symmetry is $\Z_2$, which is interpreted as an \emph{emergent} 1-form symmetry from the point of view of the above $6d\to4d$ compactification.

The fact that our predicted result for $\cL$ does not capture the true $\cL$ is not surprising as explained in the introduction. As discussed there, the predicted $\cL$ is guaranteed to match the true $\cL$ only when a non-trivial topological twist is performed on $C_g$. When no non-trivial topological twist is needed, the predicted $\cL$ is only expected to be a subgroup of the true $\cL$. In the presence of a non-trivial topological twist, the set of BPS particles would be protected as we take the limit of zero area. When there is no topological twist, the set may not be protected, as we saw in the example above where a $4d$ BPS particle (descending from the $5d$ BPS instanton particle) was decoupled in the limit of zero area. 

\subsection{Including Closed $S_3$ Twist-lines}

\begin{figure}
\centering
\begin{tikzpicture} [scale=1.9]
\draw[thick] (-1,-0.5) -- (-1,0.6);
\node at (-0.3,0.1) {$\longrightarrow$};
\node at (-1,-0.7) {$ab$};
\begin{scope}[shift={(1.3,0)}]
\draw[thick] (-1,-0.5) -- (-1,0.6);
\node at (-1,-0.7) {$b$};
\end{scope}
\begin{scope}[shift={(1.7,0)}]
\draw[thick] (-1,-0.5) -- (-1,0.6);
\node at (-1,-0.7) {$a$};
\end{scope}
\draw[thick] (0.65,0.05) -- (0.7,0.1) -- (0.75,0.05);
\begin{scope}[shift={(3.3,0.05)}]
\draw[thick] (-1,-0.5) -- (-1,0.6);
\node at (-0.3,0.1) {$\longrightarrow$};
\node at (-1,-0.7) {$a^2b$};
\begin{scope}[shift={(1.3,0)}]
\draw[thick] (-1,-0.5) -- (-1,0.6);
\node at (-1,-0.7) {$b$};
\end{scope}
\begin{scope}[shift={(1.7,0)}]
\draw[thick] (-1,-0.5) -- (-1,0.6);
\node at (-1,-0.7) {$a^2$};
\end{scope}
\draw[thick] (0.65,0.05) -- (0.7,0.1) -- (0.75,0.05);
\end{scope}
\node at (1.55,0.1) {;};
\end{tikzpicture}
\caption{Resolving various $S_3$ lines into $a$-lines and $b$-lines.}
\label{fig:resab}
\end{figure}
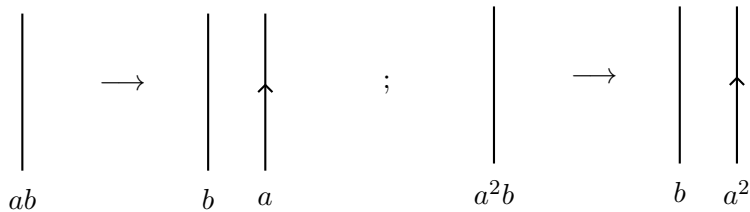

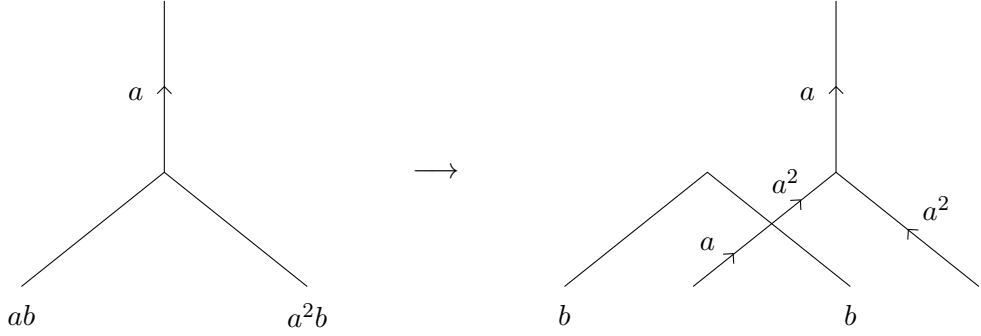
\begin{figure}
\centering
\begin{tikzpicture} [scale=1.9]
\draw (-1,-0.1) -- (0,0.7) -- (1,-0.1);
\draw (0,1.9) -- (0,0.7);
\node at (-1,-0.3) {$ab$};
\node at (1,-0.3) {$a^2b$};
\draw (-0.05,1.25) -- (0,1.3) -- (0.05,1.25);
\node at (-0.2,1.25) {$a$};
\node at (1.9,0.7) {$\longrightarrow$};
\begin{scope}[shift={(3.8,0)}]
\draw (-1,-0.1) -- (0,0.7) -- (1,-0.1);
\node at (-1,-0.3) {$b$};
\node at (1,-0.3) {$b$};
\end{scope}
\begin{scope}[shift={(4.7,0)}]
\draw (-1,-0.1) -- (0,0.7) -- (1,-0.1);
\draw (0,1.9) -- (0,0.7);
\node at (-0.9,0.2) {$a$};
\node at (0.7022,0.4493) {$a^2$};
\draw (-0.05,1.25) -- (0,1.3) -- (0.05,1.25);
\node at (-0.2,1.25) {$a$};
\end{scope}
\begin{scope}[shift={(6.0759,-0.6665)},rotate=42]
\draw (-0.05,1.25) -- (0,1.3) -- (0.05,1.25);
\end{scope}
\begin{scope}[shift={(3.1106,-0.8358)},rotate=-42]
\draw (-0.05,1.25) -- (0,1.3) -- (0.05,1.25);
\end{scope}
\begin{scope}[shift={(3.5804,-0.4595)},rotate=-42]
\draw (-0.05,1.25) -- (0,1.3) -- (0.05,1.25);
\end{scope}
\node at (4.3473,0.6378) {$a^2$};
\end{tikzpicture}
\caption{An example of resolving a trivalent $S_3$ vertex into an $a$-vertex and a $b$-vertex. Notice that two $b$ lines meet to form a trivial line (since $b^2=1$), which has not been displayed. The vertex formed by $b$ lines can now be smoothened out.}
\label{fig:resv}
\end{figure}

An arbitrary $S_3$ twist on $C_g$ can be manufactured by combining $a$ and $b$ twist lines, which are two elements of orders three and two respectively inside $S_3$ (see section \ref{6d}).  An arbitrary $S_3$ twist is described as a trivalent network of topological lines valued in $S_3$ obeying group composition law. One can separate the $a$-dependent part out of each edge in this network. That is, an edge carrying $ab$ can be separated into $b$ and $a$, and an edge carrying $a^2b$ can be separated into $b$ and $a^2$, while an edge carrying either of $1,a,a^2,b$ is left alone without any decomposition (See figure \ref{fig:resab}). Each trivalent vertex is similarly decomposed into a vertex for $b$ lines and a vertex for $a,a^2$ lines. To decompose the vertices, we have to sometimes cross an $a$ or $a^2$ line across a $b$ line. Such a crossing transforms $a$ to $a^2$ and $a^2$ to $a$. See figure \ref{fig:resv}. After decomposing the vertices, the original network has been decomposed into a network of $b$ lines, with a network of $a,a^2$ lines placed on top of the network of $b$ lines such that any time a line carrying $a$ or $a^2$ crosses a $b$ line, it is transformed to an $a^2$ or $a$ line respectively.

The network of $b$ lines can be smoothened, and hence is an element $L\in H_1(C_g,\Z_2)$ along which we wrap $b$ lines. In the previous subsection, we have seen that we can always choose $L=0$ or $L=B_1$. If we choose $L=0$, then we can represent the network of $a$ lines as an element $\ell\in H_1(C_g,\Z_3)$ with the dual element being $\wh{\ell}\in H^1(C_g,\Z_3)$. We can perform $Sp(2g,\Z)$ transformations on cycles $A_i,B_i$ to obtain a frame such that $\wh{\ell}(A_1)\in\{0,1\}$, $\wh{\ell}(B_i)=0$ for all $i$, and $\wh{\ell}(A_i)=0$ for all $i\neq 1$. If $\wh{\ell}(A_1)=0$, then we are back in the completely untwisted case discussed earlier. If $\wh{\ell}(A_1)=1$, then we have an $a$ line wrapping $B_1$. We can see that there is no element of $\wh{Z}$ that can wrap $A_1$. On the other hand, any element of $\wh{Z}$ which wraps $B_1$ can be identified with the trivial element of $\wh{Z}$ due to the action of $a$ twist line. Thus, in this case, an absolute $4d$ $\cN=2$ theory is chosen by
\be
\Lambda\subset\cL\simeq\wh{Z}^{g-1}_A \times \wh{Z}^{g-1}_B
\ee
with $\wh{Z}\simeq\Z_2\times\Z_2$. The 1-form symmetry is identified with $\wh{\Lambda}$.

Now, let us choose $L=B_1$. Since there is only a single $b$ line, it is not possible to close an $a$ line that crosses $b$ line. Using this fact, one can argue that the only possible network of $a,a^2$ lines can be represented as an element $\ell\in H_1(C_g,\Z_3)$ which does not wrap $A_1$. Let the dual element be $\wh{\ell}\in H^1(C_g,\Z_3)$, which has the property that $\wh{\ell}(B_1)=0$. Now we can perform $Sp(2g-2,\Z)$ transformations on cycles $A_i,B_i$ for $i\neq 1$ to obtain a frame such that $\wh{\ell}(A_i)\in\{0,1\}$ for $i=1,2$, $\wh{\ell}(B_i)=0$ for all $i$, and $\wh{\ell}(A_i)=0$ for all $i\neq 1,2$. If $\wh{\ell}(A_1)=1$, then in total we have the $ab$ line wrapping $B_1$ (where we are representing $S_3$ as a multiplicative group). But since $ab$ is in the same conjugacy class as $b$, we can replace $ab$ wrapping $B_1$ by $b$ wrapping $B_1$ by performing gauge transformation inside $S_3$. Thus, we can always ensure that $\wh{\ell}(A_1)=0$. Now if $\wh{\ell}(A_2)=0$, then we are back in the $\Z_2$ twisted case discussed before. 

\begin{figure}
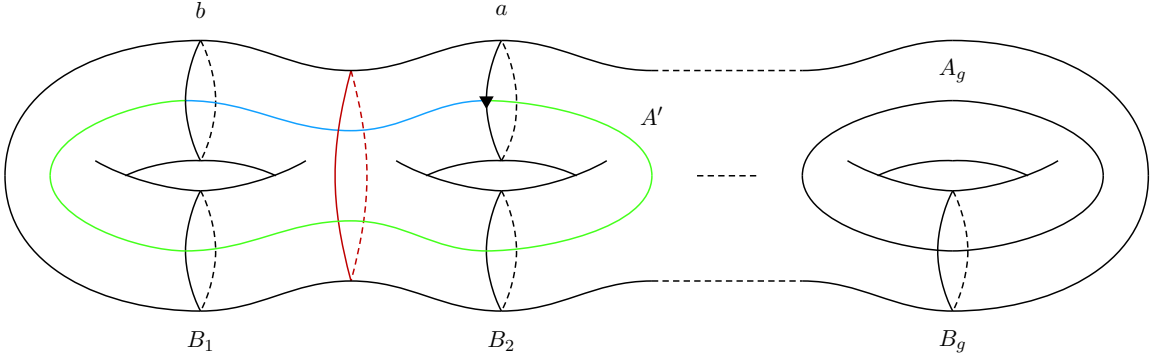

\centering
\scalebox{.8}{
}
\caption{A Riemann surface of genus $g$ with a closed $\Z_2$ twist line $b$ wrapped along the $B_1$ cycle and a closed $\Z_3$ twist line $a$ wrapped along the $B_2$ cycle. The cycle $A'$ (which is homologically equivalent to $A_1+A_2$) has been divided into two sub-segments, denoted respectively by green and blue. The color is changed as $A'$ crosses a twist line, indicating that an element of $\wh{Z}$ wrapped along one sub-segment is in general different from the element of $\wh{Z}$ wrapped along the other sub-segment, due to the action of outer-automorphism associated to the twist line.}
\label{fig:S3Twist}
\end{figure}

The only new case therefore is when $\wh{\ell}(A_1)=0$ and $\wh{\ell}(A_2)=1$, which has been represented in figure \ref{fig:S3Twist}. Consider the impact of twist lines on the elements of $\wh{Z}$ wrapping B-cycles first. Notice that $s$ wrapped along the red cycle can be identified with $v(B_1)$ (i.e. $v$ wrapped along the cycle $B_1$) by moving it to the left, and with $c(B_2)$ by moving it to the right. See figure \ref{fig:S3Twist}. Thus $v(B_1)=c(B_2)$. Similarly, $c$ wrapped along the red cycle can be identified with $v(B_1)$ by moving it to the left, and with $v(B_2)$ by moving it to the right. See figure \ref{fig:S3Twist}. Thus $v(B_1)=c(B_2)=v(B_2)$ and we deduce that the B-cycles give rise to a group
\be\label{z3tb}
\text{Proj}(\wh{Z};a,b)\simeq\Z_2\times\Z_2
\ee
generated by $s(B_1),c(B_2)$. Now, consider the impact of twist lines on the elements of $\wh{Z}$ wrapping A-cycles. We can have $v(A_1)$, but nothing can wrap $A_2$ alone. Consider the cycle $A'=A_1+A_2$ which is denoted as partly blue and partly green in figure \ref{fig:S3Twist}. We can wrap $s$ along the blue sub-segment of $A'$, and $c$ along the green sub-segment of $A'$. This configuration is consistent with the twist lines along $B_1$ and $B_2$. We label the $4d$ line operator obtained via this configuration as $s(A')$. Thus, the A-cycles give rise to the group
\be\label{z3ta}
\text{Inv}(\wh{Z};a,b)\simeq\Z_2\times\Z_2
\ee
generated by $v(A_1),s(A')$. It can be easily seen that the pairs of dual elements in $\text{Inv}(\wh{Z};a,b)\times\text{Proj}(\wh{Z};a,b)$ are $\{v(A_1),s(B_1)\},\{s(A'),c(B_2)\}$. Thus, an absolute $4d$ $\cN=2$ theory is chosen by
\be
\Lambda\subset \text{Inv}(\wh{Z};a,b) \times \text{Proj}(\wh{Z};a,b)\times \wh{Z}^{g-2}_A \times \wh{Z}^{g-2}_B
\ee
with $\text{Inv}(\wh{Z};a,b)$ and $\text{Proj}(\wh{Z};a,b)$ given by (\ref{z3ta}) and (\ref{z3tb}) respectively. The 1-form symmetry of this absolute $4d$ $\cN=2$ theory is identified with $\wh{\Lambda}$.

\section{Compactifications with Regular Punctures}\label{w/p}

\begin{figure}
\centering
\scalebox{1}{
\begin{tikzpicture}
	\begin{pgfonlayer}{nodelayer}
		\node [style=none] (0) at (1.5, 0) {};
		\node [style=none] (1) at (4.5, 0) {};
		\node [style=NodeCross] (2) at (-4.5, 0) {};
		\node [style=none] (3) at (3, 0) {};
		\node [style=NodeCross] (4) at (1.5, 0) {};
		\node [style=ltriangle] (5) at (3, 0) {};
		\node [style=none] (6) at (3, 0.5) {};
		\node [style=none] (7) at (3, 0.5) {$t$};
		\node [style=none] (8) at (3, -1.5) {Twisted Puncture};
		\node [style=none] (10) at (-4.5, -1.5) {Untwisted Puncture};
	\end{pgfonlayer}
	\begin{pgfonlayer}{edgelayer}
		\draw [style=ThickLine] (0.center) to (1.center);
	\end{pgfonlayer}
\end{tikzpicture}}
\caption{A twisted puncture lives at the end of a non-trivial twist line $t$, while an untwisted puncture does not live at the end of a non-trivial twist line.}
\label{fig:UntwistedTwistedPuncture}
\end{figure}
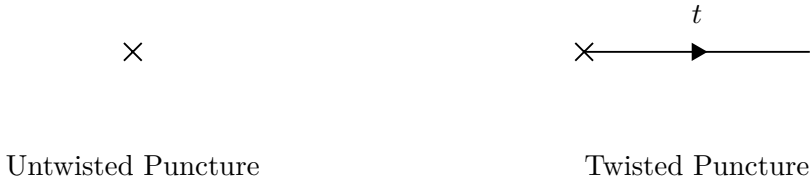

Regular punctures are a special set of punctures defined by the condition that the Hitchin field has (at most) a simple pole at the location of the puncture. These punctures can be either \emph{untwisted} or \emph{twisted}. Twisted regular punctures arise at the ends of twist lines, and hence the Hitchin field transforms by the action of the corresponding outer-automorphism as one encircles a twisted regular puncture. On the other hand, untwisted punctures do \emph{not} live at the ends of non-trivial twist lines, and correpondingly the Hitchin field does not pick up the action of any non-trivial outer automorphism as one encircles an untwisted regular puncture. See figure \ref{fig:UntwistedTwistedPuncture}.

Moreover, we need to consider a rather small, special subset of regular punctures separately. The punctures in this subset are referred to as \emph{atypical} punctures. In the presence of atypical regular punctures, the number of simple factors in the gauge algebra arising in a degeneration limit of the Riemann surface is not equal to the dimension of the moduli space of the Riemann surface \cite{Chacaltana:2012ch,Chacaltana:2013oka,Chacaltana:2016shw} (see also \cite{Tachikawa:2009rb}). We call a regular puncture which is not atypical as a \emph{typical} puncture. An atypical regular puncture can be resolved into some number of typical regular punctures. Throughout this section until subsection \ref{AP}, a regular puncture always refers to a \emph{typical} regular puncture.

In this section, we consider compactifications of $6d$ $\cN=(2,0)$ theories on a Riemann surface $C_g$ with an arbitrary number of (untwisted and twisted) regular punctures, and an arbitrary number of closed twist lines (which do not have end-points).

\subsection{Untwisted Regular Punctures}
Let $\cL$ be the set of $4d$ line operators (modulo screening) when a $(2,0)$ theory is compactified on a Riemann surface $C_g$ without any punctures, but possibly in the presence of closed twist lines. The set $\cL$ (and Dirac pairing on it) was determined in the last few subsections. Now, insert $n$ regular untwisted punctures on $C_g$. We propose that the set of $4d$ line operators modulo flavor charges (and screening) can again be identified with $\cL$. Moreover, an absolute $4d$ $\cN=2$ theory is obtained by choosing a maximal isotropic subgroup
\be
\Lambda\subset\cL
\ee
and the 1-form symmetry of such an absolute $4d$ $\cN=2$ theory can be identified with $\wh{\Lambda}$. In other words, regular untwisted punctures turn out to be irrelevant for the considerations of this paper. In the rest of this section, we substantiate this proposal by studying some examples.

\vspace{6mm}

\ni\ubf{Sphere with 4 regular untwisted punctures}: As a few examples, we can obtain the following $4d$ $\cN=2$ gauge theories by compactifying $(2,0)$ theories on a sphere with 4 regular untwisted punctures\footnote{The notation $\fg_i+\sum n_i \mathsf{R}_i$ denotes a $4d$ $\cN=2$ gauge theory with gauge algebra $\fg$ along with $n_i$ full hypers in irrep $\mathsf{R}_i$. If $n_i$ is half-integral, it means that there is an additional half-hyper in $\mathsf{R}_i$ along with $\lfloor n_i\rfloor$ number of full hypers in $\mathsf{R}_i$. $\F$ denotes fundamental irrep for $\su(n)$ and $\sp(n)$, and vector irrep for $\so(n)$. $\S$ denotes spinor irreps for $\so(n)$ and $\C$ denotes co-spinor irrep for $\so(2n)$. $\Lambda^2$ denotes 2-index antisymmetric irrep for $\su(n)$ and $\sp(n)$. See also appendix \ref{app:Not}.}:
\bit
\item $\su(n)+2n\F$ by compactifying $A_{n-1}$ $(2,0)$ theory.
\item $\so(8)+2\F+2\S+2\C,~\so(8)+4\S+2\C,~\so(8)+4\S+\C+\F$ by compactifying $D_{4}$ $(2,0)$ theory \cite{Chacaltana:2013dsj}.
\item $\so(9)+3\S+\F,~\so(10)+4\S,~\so(10)+2\S+4\F$ by compactifying $D_{5}$ $(2,0)$ theory \cite{Chacaltana:2013dsj}.
\item $\so(11)+\S+5\F,~\so(11)+\frac32\S+3\F,~\so(12)+\S+\half\C+4\F,~\so(12)+\S+6\F,~\so(12)+\frac32\S+\half\C+2\F$ by compactifying $D_{6}$ $(2,0)$ theory \cite{Chacaltana:2013dsj}.
\item $\su(4)+2\L^2+4\F,~\sp(2)+6\F$ by compactifying $A_{3}$ $(2,0)$ theory \cite{Chacaltana:2010ks}.
\eit
For this case $\cL$ is trivial, which is what is expected from the $4d$ gauge theory description as it can be checked that the line operators (modulo screening and flavor charges) form a trivial set in all of the above gauge theories. Consequently, the 1-form symmetry is also trivial for all of these theories, and the gauge group must be the simply connected one.

\vspace{6mm}

\ni\ubf{Torus with 1 regular untwisted puncture and twisted line}: We can obtain the following $4d$ $\cN=2$ gauge theories by compactifying $(2,0)$ theories on a torus with 1 regular untwisted puncture and a twisted line wrapped along a non-trivial cycle\footnote{$\S^2$ denotes the 2-index symmetric representation of $\su(n)$. See also appendix \ref{app:Not}.}:
\bit
\item $\su(2n)+\S^2+\L^2$ by compactifying $A_{2n-1}$ $(2,0)$ theory.
\item $\su(2n+1)+\S^2+\L^2$ by compactifying $A_{2n}$ $(2,0)$ theory.
\eit
In the former case, we have
\be\label{pred}
\cL\simeq\Z_2\times\Z_2 \,,
\ee
which can be matched with the $4d$ gauge theory expectation. For a pure $\su(2n)$ gauge theory, the set of Wilson lines (modulo screening) is $\Z_{2n}$ with generator $W$ being the Wilson line in fundamental rep of $\su(2n)$. The set of 't Hooft lines (modulo screening) is also $\Z_{2n}$ with generator $H$. The Dirac pairing between $W$ and $H$ is $\langle W,H\rangle=\frac{1}{2n}$. Now we add in the matter. The hypermultiplets in $\S^2$ and $\L^2$ screen $2W$, and thus the set of Wilson lines (modulo screening and flavor charges) can be identified with $\Z_2$, generated by $W$. On the other hand, the 't Hooft lines must be mutually local with $2W$, and hence the set of 't Hooft lines (modulo screening and flavor charges) can be identified with $\Z_2$, generated by $nH$. Thus, we verify the prediction (\ref{pred}). Choosing the polarization $\Lambda$ to be the $\Z_2$ generated by $W$ leads to gauge group $SU(2n)$. Choosing $\Lambda$ to be the $\Z_2$ generated by $nH$ or $W+nH$ leads to gauge group $SU(2n)/\Z_2$ with discrete theta parameter turned off or on respectively. In all these cases, the 1-form symmetry is 
\be
\wh{\Lambda}\simeq\Z_2 \,.
\ee

In the latter case, $\cL$ is trivial. Correspondingly, the set of line operators in the gauge theory (modulo screening and flavor charges) is trivial. The set of Wilson lines is trivial because $2W$ is a generator of $\Z_{2n+1}$, and the set of 't Hooft lines is trivial because they need to be mutually local with $W$ (as $W$ is screened). There is no 1-form symmetry, and the gauge group must be the simply connected $SU(2n+1)$.

\vspace{6mm}

\ni\ubf{Torus with $k$ regular untwisted punctures}: $4d$ $\cN=2$ $\su(n)^k$ necklace quiver can be obtained by compactifying $A_{n-1}$ $(2,0)$ theory on a torus with $k$ regular untwisted punctures. In this case,
\be
\cL\simeq\wh{Z}_A\times\wh{Z}_B\simeq\Z^A_n\times\Z^B_n \,,
\ee
which can be verified from the $4d$ gauge theory description. For example, choosing all gauge groups to be $SU(n)$ corresponds to choosing one of the two $\Z_n$ factors as the polarization. The 1-form symmetry is then predicted to be
\be
\wh{\Lambda}\simeq\Z_n\,,
\ee
which can be identified as the diagonal subgroup of the $\Z^k_n$ center of the gauge group $SU(n)^k$.

\vspace{6mm}

\ni\ubf{$C_g$ with $n$ regular untwisted punctures}: Consider compactifying $A_1$ $(2,0)$ theory on $C_g$ in the presence of $n$ regular untwisted punctures \cite{Gaiotto:2009we}. According to our proposal, we predict
\be
\cL\simeq\Z_2^g\times\Z_2^g\,.
\ee
There are a number of degeneration limits which lead to a variety of S-dual weakly-coupled $4d$ conformal gauge theories. The predicted answer for $\cL$ and the pairing on it can be verified from the point of view of any of these $4d$ gauge theories. For example, one such degeneration limit (which exists for $n\ge2$) leads to the following $4d$ gauge theory
\be
\scalebox{0.95}{
\begin{tikzpicture} [scale=1.9]
\node (v1) at (-5.6,0.4) {$\su(2)$};
\node (v2) at (-6.6,0.4) {$\su(2)$};
\draw  (v2) edge (v1);
\node (v3) at (-4.9,0.4) {$\cdots$};
\draw  (v1) edge (v3);
\node (v4) at (-4.2,0.4) {$\su(2)$};
\draw  (v3) edge (v4);
\node (v5) at (-3.2,0.4) {$\so(4)$};
\draw  (v4) edge (v5);
\node (v6) at (-6.6,1.1) {$2\F$};
\begin{scope}[shift={(-2.7,0.05)}]
\node at (-2.7,-0.1) {$n-1$};
\draw (-4.2,0.2) .. controls (-4.2,0.15) and (-4.2,0.1) .. (-4.1,0.1);
\draw (-4.1,0.1) -- (-2.8,0.1);
\draw (-2.7,0.05) .. controls (-2.7,0.1) and (-2.75,0.1) .. (-2.8,0.1);
\draw (-2.7,0.05) .. controls (-2.7,0.1) and (-2.65,0.1) .. (-2.6,0.1);
\draw (-2.6,0.1) -- (-1.3,0.1);
\draw (-1.2,0.2) .. controls (-1.2,0.15) and (-1.2,0.1) .. (-1.3,0.1);
\end{scope}
\draw  (v6) edge (v2);
\node (v7) at (-2.2,0.4) {$\su(2)$};
\node (v8) at (-1.5,0.4) {$\cdots$};
\node (v9) at (-0.8,0.4) {$\so(4)$};
\node (v10) at (0.2,0.4) {$\su(2)$};
\node (v11) at (1.2,0.4) {$\so(3)$};
\draw  (v5) edge (v7);
\draw  (v7) edge (v8);
\draw  (v8) edge (v9);
\draw  (v9) edge (v10);
\draw  (v10) edge (v11);
\begin{scope}[shift={(0.7,0.05)}]
\node at (-1.7,-0.1) {$2g-1$};
\draw (-4.2,0.2) .. controls (-4.2,0.15) and (-4.2,0.1) .. (-4.1,0.1);
\draw (-4.1,0.1) -- (-1.8,0.1);
\draw (-1.7,0.05) .. controls (-1.7,0.1) and (-1.75,0.1) .. (-1.8,0.1);
\draw (-1.7,0.05) .. controls (-1.7,0.1) and (-1.65,0.1) .. (-1.6,0.1);
\draw (-1.6,0.1) -- (0.7,0.1);
\draw (0.8,0.2) .. controls (0.8,0.15) and (0.8,0.1) .. (0.7,0.1);
\end{scope}
\node (v12) at (0.2,1.1) {$\half\F$};
\draw  (v12) edge (v10);
\end{tikzpicture}}
\ee
where an edge between two $\su(2)$ gauge algebras denotes a full hyper in bifundamental, while an between an $\su(2)$ and an $\so(n)$ gauge algebra denotes a half-hyper in bifundamental (see earlier discussion for our slightly non-standard definition of fundamental of $\so(3)$ and $\so(4)$). The edge between a node labeled $n\F$ and a node labeled $\su(2)$ denotes that the corresponding $\su(2)$ gauge algebras carries $n$ extra hypers in fundamental representation, where $n$ is allowed to be a half-integer to account for half-hypers in fundamental. Choosing a particular $\Lambda\simeq\Z_2^g\subset\cL$ corresponds to choosing all the gauge groups to be simply connected. The 1-form symmetry is predicted to be $\wh{\Lambda}\simeq\Z_2^g$ for this choice, which can be verified easily from the $4d$ gauge theory description. A $\Z_2$ factor arises as the subgroup of the center of each Spin$(n)$ (where $n=3,4$) gauge group that leaves the vector rep of Spin$(n)$ invariant.

\paragraph{\underline{Example and Comparison with $6d$ $(1,0)$ on $T^2$:}} 
The last class of example has an alternative realization in terms of a $6d$ $(1,0)$ on $T^2$ \cite{Ohmori:2015pua, Ohmori:2015pia}: For $g=1$ and $n=2$ the $A_1$ theory on $C_{1,2}$ has defect group $\mathcal{L} = \mathbb{Z}_2 \times \mathbb{Z}_2$. We can alternatively think of this as the compactification of the $6d$ (1,0) theory that is the $SU(2)-SU(2)$ conformal matter theory of rank 2, i.e. 2 M5-branes probing $\mathbb{C}^2/\mathbb{Z}_2$. The 6d theory has a tensor branch geometry, which has two non-compact curves, with $SU(2)$ singularities, sandwiching a $(-2)$-curve, with $SU(2)$ gauge group. The defect group given by $\mathbb{Z}_2$, and the dimensional reduction of this on $T^2$, results in $\mathcal{L}_A = \mathcal{L}_B = \mathbb{Z}_2$. 
More generally, 2 M5-branes probing a $\mathbb{Z}_k$ singularity results in a `hybrid' class S theory, where an $A_1$-trinion is glued to an $A_{k-1}$ one (see (2.6) in \cite{Ohmori:2015pia}). The tensor branch-geometry changes simply to $SU(k)$ groups both on the non-compact curves as well as on the $(-2)$-curve, thus leaving the defect group, and the expected 1-form symmetry unchanged. 

\begin{figure}
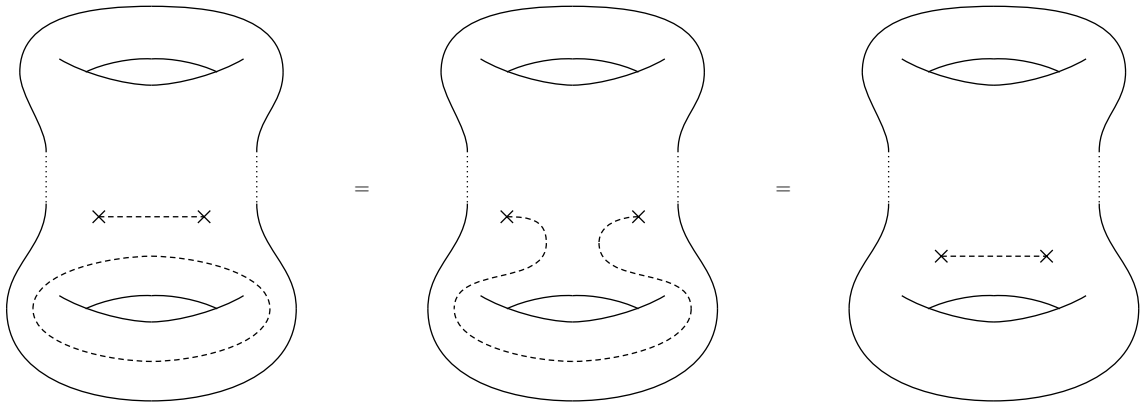

\centering
\scalebox{.7}{

}
\caption{A configuration involving a closed $\Z_2$ twist line and an open $\Z_2$ twist line is topologically equivalent to a configuration involving on an open $\Z_2$ twist line.}
\label{fig:CombiningTwistLines}
\end{figure}

\subsection{$\Z_2$-twisted Regular Punctures}

In this subsection we either consider those $\fg$ for with the outer-automorphism group is $\Z_2$, or the case $\fg=D_4$ with twist lines valued only in the $\Z_2$ subgroup of the $S_3$ outer-automorphism group generated by the element $b$ (see section \ref{6d}).

Consider a $(2,0)$ theory compactified on $C_g$ with $\Z_2$ twist lines, in the presence of both untwisted\footnote{From this point onward, the reader should assume that an arbitrary number of untwisted regular punctures are always present. We do not mention them in what follows since they do not enter in the computation of $\cL$ and 1-form symmetry $\wh{\Lambda}$.} and twisted regular punctures. Twisted regular punctures are the regular punctures that appear at the ends of open $\Z_2$ twist lines. First thing to note is that if we have an open twist line (i.e. a twist line with two end points), then we can always remove all the closed $\Z_2$ twist lines, since combining a closed twist line with an open twist line results in a single open twist line, shown in figure \ref{fig:CombiningTwistLines}.

\begin{figure}
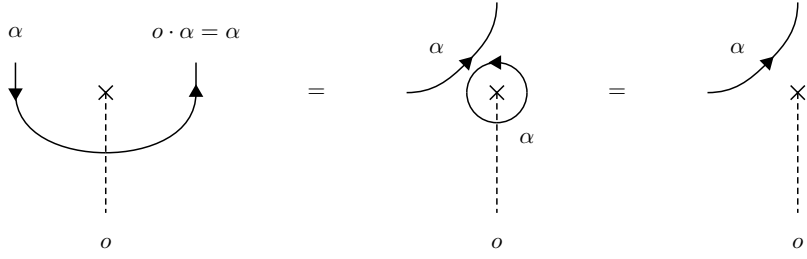

\centering
\scalebox{.8}{

}
\caption{An element $\alpha\in\wh{Z}$ with the property $o\cdot\alpha=o$ can be moved across a regular twisted puncture living at the end of an $o$ twist line. Left: $\alpha$ crossing a $\Z_2$ twist line $o$ ending on the regular puncture. Center: A topological deformation creates a loop of $\alpha$ around the regular puncture, and an $\alpha$ line passing nearby that does not cross the twist line. Right: The loop of $\alpha$ surrounding the regular twisted puncture can be collapsed using our main proposal.}
\label{fig:PullOverMove}
\end{figure}

Recall that in the last subsection we proposed that any element of $\wh{Z}$ that can be inserted on a loop surrounding an untwisted regular puncture does not contribute to the set of $4d$ line defects (modulo flavor charges). Similarly, we propose that any element of $\wh{Z}$ that can be inserted on a loop surrounding a twisted regular puncture does not contribute to the set of $4d$ line defects (modulo flavor charges) either. However, notice that the only elements of $\wh{Z}$ that can be inserted along a loop surrounding a single twisted regular puncture are those that are left invariant by the $\Z_2$ outer-automorphism action. The set of such invariant elements was denoted by $\text{Inv}(\wh{Z}, o)$ earlier. So, according to this proposal, only the elements in the subset $\text{Inv}(\wh{Z}, o)$ can be moved across a twisted regular puncture, while the elements not in the subset $\text{Inv}(\wh{Z}, o)$ cannot be moved across a twisted regular puncture. See figure \ref{fig:PullOverMove}.

\begin{figure}
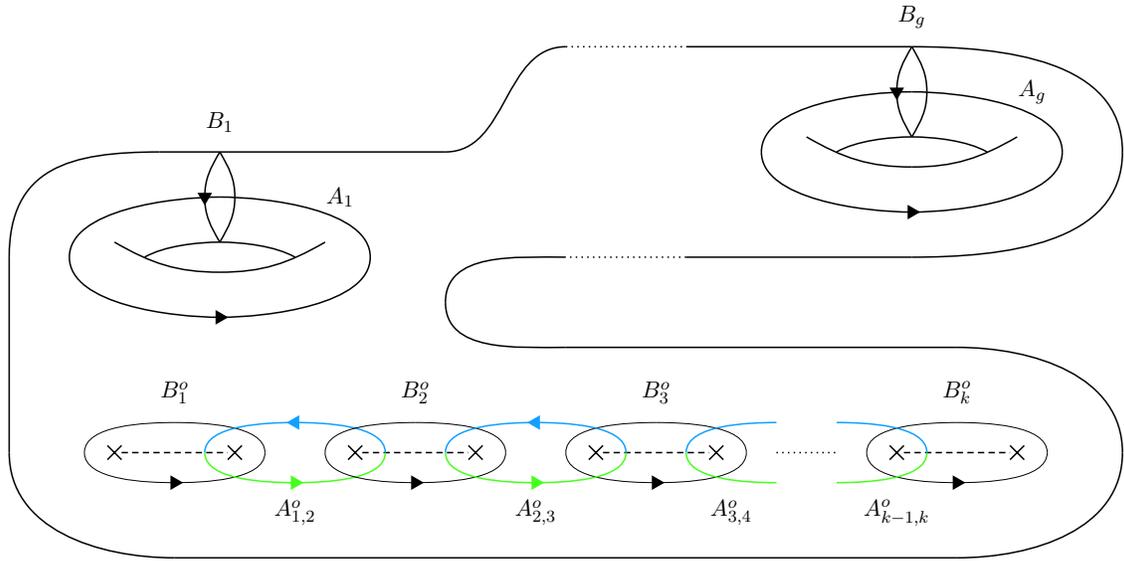

\centering
\scalebox{.8}{

}
\caption{A Riemann surface with $2k$ $\Z_2$ twisted regular punctures and $k$ open twist lines of type $o$, all collected in a corner of the genus $g$ Riemann surface. From this point onward, an arbitrary number of untwisted regular punctures are always present, but are never displayed. Further we fix cycles associated with punctures and twist lines to be oriented counterclockwise and omit orientations in future pictures. Similarly we fix the orientation of the $A,B$ cycles of the Riemann surface as shown above going forward.}
\label{fig:CornerCollect}
\end{figure}

Now, in order to determine the set of $4d$ line defects, we collect all the twisted regular punctures in one corner of $C_g$ as shown in figure \ref{fig:CornerCollect}. From our above proposal we deduce that the set of line operators $\cL^B_i$ originating from elements of $\wh{Z}$ wrapping the cycle $B^o_i$ can be identified with $\wh{Z}/\text{Inv}(\wh{Z}, o)$. Furthermore, we can parametrize the set of line operators $\cL^A_{i,i+1}$ originating from the cycle $A^o_{i,i+1}$ with the elements of $\wh{Z}$ inserted along the green sub-segment of $A^o_{i,i+1}$. If we insert an element $\alpha$ of the subgroup $\text{Inv}(\wh{Z}, o)$ along the green sub-segment, then the element inserted along the blue sub-segment is also $\alpha$. This configuration can be moved across the twisted regular punctures and hence trivialized. Thus, $\cL^A_{i,i+1}$ can also be identified with $\wh{Z}/\text{Inv}(\wh{Z}, o)$. Finally, notice that any element wrapped along $\sum_i B^o_i$ can be unwrapped on the other side of $C_g$, and hence any element wrapped along $B^o_n$ can be written in terms of elements wrapped along $B^o_i$ for $1\le i\le k-1$.

Thus, the set of $4d$ line operators (modulo screening and flavor charges) $\cL$ can be written as
\be\label{LZ2}
\cL\simeq\prod_{i=1}^{k-1}\cL^A_{i,i+1}\times\prod_{i=1}^{k-1}\cL^B_i\times\wh{Z}^g_A\times\wh{Z}^g_B \,.
\ee
For $\fg=A_{2n-1}$, we have
\be \label{eq:LineOp1}
\cL^A_{i,i+1}\simeq\cL^B_i\simeq\Z_n \,.
\ee
We choose $f$ wrapped along $B^o_i$ as the generator $g^B_i$ of $\cL^B_i$, and the element obtained by wrapping $f$ along the green sub-segment of $A^o_{i,i+1}$ to be the generator $g^A_{i,i+1}$ of $\cL^A_{i,i+1}$. Then, the non-trivial pairings\footnote{The pairing is a product of an intersection number and the value of the bihomomorphism \eqref{eq:BiHom}. Intersections are taken to be positive if complementing the direction of the first and second argument by a vector pointing outward from the page results in a right handed basis.} mod 1 are
\begin{align}
\langle g^A_{i,i+1},g^B_i\rangle&=\frac1n\\
\langle g^A_{i,i+1},g^B_{i+1}\rangle&=-\frac1n
\end{align}
along with the previously discussed pairing on $\wh{Z}^g_A\times\wh{Z}^g_B$.

For $\fg=A_{2n}$, we have
\be\label{eq:LineOp2}
\cL^A_{i,i+1}\simeq\cL^B_i\simeq\Z_{2n+1}
\ee
We choose $f$ wrapped along $B^o_i$ as the generator $g^B_i$ of $\cL^B_i$, and the element obtained by wrapping $f$ along the green sub-segment of $A^o_{i,i+1}$ to be the generator $g^A_{i,i+1}$ of $\cL^A_{i,i+1}$. Then, the non-trivial pairings are
\begin{align}
\langle g^A_{i,i+1},g^B_i\rangle&=\frac2{2n+1}\\
\langle g^A_{i,i+1},g^B_{i+1}\rangle&=-\frac2{2n+1} \,.
\end{align}
For $\fg=D_{n}$, we have
\be\label{LidD}
\cL^A_{i,i+1}\simeq\cL^B_i\simeq\Z_{2} \,.
\ee
We choose $s$ wrapped along $B^o_i$ as the generator $g^B_i$ of $\cL^B_i$, and the element obtained by wrapping $s$ along the green sub-segment of $A^o_{i,i+1}$ to be the generator $g^A_{i,i+1}$ of $\cL^A_{i,i+1}$. Then, the non-trivial pairings are
\begin{align}
\langle g^A_{i,i+1},g^B_i\rangle&=\half\label{LpD1}\\
\langle g^A_{i,i+1},g^B_{i+1}\rangle&=\half\label{LpD2} \,.
\end{align}
For $\fg=E_{6}$, we have
\be\label{eq:LineOp4}
\cL^A_{i,i+1}\simeq\cL^B_i\simeq\Z_{3} \,.
\ee
We choose $f$ wrapped along $B^o_i$ as the generator $g^B_i$ of $\cL^B_i$, and the element obtained by wrapping $f$ along the green sub-segment of $A^o_{i,i+1}$ to be the generator $g^A_{i,i+1}$ of $\cL^A_{i,i+1}$. Then, the non-trivial pairings are
\begin{align}
\langle g^A_{i,i+1},g^B_i\rangle&=\frac13\\
\langle g^A_{i,i+1},g^B_{i+1}\rangle&=-\frac13 \,.
\end{align}
With these pairings, an absolute $4d$ $\cN=2$ theory is chosen by a maximal isotropic subgroup
\be
\Lambda\subset\cL\simeq\prod_{i=1}^{k-1}\cL^A_{i,i+1}\times\prod_{i=1}^{k-1}\cL^B_i\times\wh{Z}^g_A\times\wh{Z}^g_B
\ee
The $4d$ theory carries a 1-form symmetry $\wh{\Lambda}$. In the rest of this subsection, we substantiate our proposal by discussing a few Lagrangian examples.

\vspace{6mm}

\ni\ubf{Sphere with 2 regular twisted punctures}: The following $4d$ $\cN=2$ theories can be produced by compactifying $(2,0)$ theories on a sphere with 2 regular twisted punctures and 2 regular untwisted punctures:
\bit
\item $\sp(n-1)+2n\F$ by compactifying $D_n$ $(2,0)$ theory.
\item $\sp(3)+3\F+\L^3,~\sp(3)+\frac{11}2\F+\half\L^3,~\so(8)+8\F+\S,~\so(7)+4\F+\S$ by compactifying $D_4$ $(2,0)$ theory \cite{Chacaltana:2013oka}.
\item $\so(10)+\S+6\F,~\so(9)+\S+5\F$ by compactifying $D_5$ $(2,0)$ theory \cite{Chacaltana:2013oka}.
\item $\so(12)+\half\S+8\F,~\so(11)+\half\S+7\F$ by compactifying $D_6$ $(2,0)$ theory \cite{Chacaltana:2013oka}.
\item $\so(13)+\half\S+7\F$ by compactifying $D_7$ $(2,0)$ theory \cite{Chacaltana:2013oka}.
\item $\su(4)+3\L^2+2\F,~\sp(2)+2\L^2+2\F$ by compactifying $A_3$ $(2,0)$ theory \cite{Chacaltana:2012ch}.
\eit
In such a compactification, our proposal predicts that $\cL$ is trivial, which can be verified by computing the set of line operators (modulo screening and flavor charges) in all of the above gauge theories. Hence, the 1-form symmetry is trivial and the gauge group in all these examples must be the simply connected one.

\vspace{6mm}

\ni\ubf{Sphere with 4 regular twisted punctures}: The following $4d$ $\cN=2$ theories can be produced by compactifying $(2,0)$ theories of type $\fg\neq D_4$ on a sphere with 4 regular twisted punctures:
\bit
\item $\so(2n)+(2n-2)\F,~\so(2n-1)+(2n-3)\F$ by compactifying $D_n$ $(2,0)$ theory.
\item $\su(4)+4\L^2,~\sp(2)+3\L^2$ by compactifying $A_3$ $(2,0)$ theory \cite{Chacaltana:2012ch}.
\eit
In both of these cases we have
\be
\cL\simeq\cL^A_{1,2}\times\cL^B_1\simeq\Z_2\times\Z_2
\ee
with the generators $g^A_{1,2}$ and $g^B_1$ having the pairing
\be
\langle g^A_{1,2},g^B_1\rangle=\half \,,
\ee
which can be verified from the perspective of Wilson-'t Hooft line operators in all of the above mentioned $4d$ gauge theories. The 1-form symmetry $\wh{\Lambda}$ for any choice of $\Lambda\subset\cL$ is
\be
\wh{\Lambda}\simeq\Z_2\,,
\ee
which can also be easily verified. For any consistent choice of gauge group and discrete theta parameters in the above gauge theories, the 1-form symmetry of the gauge theory is $\Z_2$.

Let us consider another example, which is of the $4d$ $\cN=2$ quiver gauge theory
\be
\begin{tikzpicture} [scale=1.9]
\node (v1) at (-5.6,0.4) {$\su(2n)$};
\node (v2) at (-6.6,0.4) {$\su(n)$};
\draw  (v2) edge (v1);
\node (v3) at (-4.9,0.4) {$\cdots$};
\draw  (v1) edge (v3);
\node (v4) at (-4.2,0.4) {$\su(2n)$};
\draw  (v3) edge (v4);
\node (v5) at (-3.2,0.4) {$\su(n)$};
\draw  (v4) edge (v5);
\node (v6) at (-5.6,1.1) {$\su(n)$};
\begin{scope}[shift={(-2.7,0.05)}]
\node at (-2.2,-0.1) {$k$};
\draw (-3.2,0.2) .. controls (-3.2,0.15) and (-3.2,0.1) .. (-3.1,0.1);
\draw (-3.1,0.1) -- (-2.3,0.1);
\draw (-2.2,0.05) .. controls (-2.2,0.1) and (-2.25,0.1) .. (-2.3,0.1);
\draw (-2.2,0.05) .. controls (-2.2,0.1) and (-2.15,0.1) .. (-2.1,0.1);
\draw (-2.1,0.1) -- (-1.3,0.1);
\draw (-1.2,0.2) .. controls (-1.2,0.15) and (-1.2,0.1) .. (-1.3,0.1);
\end{scope}
\node (v7) at (-4.2,1.1) {$\su(n)$};
\draw  (v7) edge (v4);
\draw  (v6) edge (v1);
\end{tikzpicture}
\ee
where an edge between two nodes denotes a bifundamental hyper between the corresponding gauge algebras. This theory can be produced by compactifying $A_{2n-1}$ $(2,0)$ theory on a sphere with 4 regular twisted punctures and $k+3$ regular untwisted punctures \cite{Chacaltana:2012ch}. In this case, our proposal predicts that
\be\label{LDq}
\cL\simeq\cL^A_{1,2}\times\cL^B_1\simeq\Z_n\times\Z_n \,,
\ee
with the generators $g^A_{1,2}$ and $g^B_1$ having the pairing
\be
\langle g^A_{1,2},g^B_1\rangle=\frac1n \,,
\ee
which can be verified from the point of view of the $4d$ gauge theory as well. For example, choosing one of the $\Z_n$ factors in (\ref{LDq}) as polarization leads to the choice of simply connected gauge group for all gauge algebras involved in the $4d$ gauge theory. The 1-form symmetry
\be
\wh{\Lambda}\simeq\Z_n
\ee
can then be identified from the gauge theory viewpoint as follows. Each bifundamental hyper between two $SU(2n)$ groups, only preserves the diagonal part of the two $\Z_{2n}$ centers, while a bifundamental hyper between an $SU(n)$ group and an $SU(2n)$ group preserves the diagonal $\Z_n$ of the obvious $\Z_n\times\Z_n$ subgroup of the center $\Z_n\times\Z_{2n}$. Thus, in total, only a diagonal $\Z_n$ of the $\Z_n^{k+4}$ subgroup of the $\Z_n^4\times\Z_{2n}^k$ center of the total gauge group acts trivially on all the matter content.

\vspace{6mm}

\ni\ubf{Torus with 6 regular twisted punctures}: The $4d$ $\cN=2$ quiver
\be
\begin{tikzpicture} [scale=1.9]
\node (v1) at (-6.1,0.4) {$\so(4n+2)$};
\node (v2) at (-7.3,0.4) {$\sp(2n)$};
\draw  (v2) edge (v1);
\node (v3) at (-4.9,0.4) {$\sp(2n)$};
\draw  (v1) edge (v3);
\node (v4) at (-3.7,0.4) {$\so(4n+2)$};
\draw  (v3) edge (v4);
\node (v5) at (-2.5,0.4) {$\sp(2n)$};
\draw  (v4) edge (v5);
\node (v6) at (-4.9,1.2) {$\so(4n+2)$};
\draw  (v6) edge (v2);
\draw  (v6) edge (v5);
\end{tikzpicture}
\ee
can be constructed by compactifying $D_{2n+1}$ $(2,0)$ theory on a torus with 6 regular twisted punctures \cite{Tachikawa:2009rb}. Our proposal would predict that for this gauge theory we have
\be
\cL\simeq\cL^A_{1,2}\times\cL^A_{2,3}\times\cL^B_1\times\cL^B_2\times\wh{Z}_A\times\wh{Z}_B
\ee
with
\be
\cL^A_{1,2}\simeq\cL^A_{2,3}\simeq\cL^B_1\simeq\cL^B_2\simeq\Z_2
\ee
and
\be
\wh{Z}_A\simeq\wh{Z}_B\simeq\Z_4
\ee
The non-trivial pairings on $\cL$ are defined in terms of generators $g^A_{i,i+1},g^B_i,g_A,g_B$ of $\cL^A_{i,i+1},\cL^B_i,$ $\wh{Z}_A,\wh{Z}_B$ respectively
\be\label{pair}
\langle g^A_{1,2},g^B_1\rangle=\half,~~\langle g^A_{1,2},g^B_2\rangle=\half,~~\langle g^A_{2,3},g^B_2\rangle=\half,~~\langle g_A,g_B\rangle=\frac14 \,.
\ee
Let us reproduce this result by explicitly studying the line operators of the $4d$ gauge theory. Before accounting for the matter content, the Wilson lines for all the gauge algebra factors form the group
\be
Z_W\simeq\prod_{i=1}^3(\Z_4)_i\times\prod_{i=1}^3(\Z_2)_i \,,
\ee
where $(\Z_4)_i$ is associated to gauge algebra $\so(4n+2)_i$, and $(\Z_2)_i$ is associated to gauge algebra $\sp(2n)_i$. We choose generators $W^\so_i$ for $(\Z_4)_i$ and $W^\sp_i$ for $(\Z_2)_i$. The matter content implies that the set of Wilson lines (modulo screening) can be generated by $W^\so_1-W^\so_2,~W^\so_2-W^\so_3,W^\so_3$. The first two generators are of order two, and the last generator is of order four. Thus, the contribution of Wilson lines to the set of line operators (modulo screening and flavor charges) is
\be
\cL_W\simeq\Z_2\times\Z_2\times\Z_4
\ee
with the generators identified above. On the other hand, before accounting for the matter content, the 't Hooft lines for all the gauge algebra factors form the group
\be
Z_H\simeq\prod_{i=1}^3(\Z_4)_i\times\prod_{i=1}^3(\Z_2)_i \,,
\ee
where $(\Z_4)_i$ is associated to gauge algebra $\so(4n+2)_i$, and $(\Z_2)_i$ is associated to gauge algebra $\sp(2n)_i$. We choose generators $H^\so_i$ for $(\Z_4)_i$ and $H^\sp_i$ for $(\Z_2)_i$. The matter content requires us to choose the subset $\cL_H$ of $Z_H$ which is mutually local with the matter content. We can choose the generators for $\cL_H$ to be $2H^\so_1,2H^\so_2,~\sum_i(H^\so_i+H^\sp_i)$. The first two generators are of order two, and the last generator is of order four. Thus, the contribution of 't Hooft lines to the set of line operators (modulo screening and flavor charges) is
\be
\cL_H\simeq\Z_2\times\Z_2\times\Z_4
\ee
with the generators identified above. We thus see that clearly
\be
\cL_W\times\cL_H\simeq\cL
\ee
and the generators can be identified as
\be
g^A_{i,i+1}=W^\so_i-W^\so_{i+1},~~g^B_i=2H^\so_i,~~g_A=W^\so_3,~~g_B=\sum_i(H^\so_i+H^\sp_i) \,.
\ee
It is straightforward to check that the Dirac pairing between Wilson and 't Hooft lines reproduces the pairing (\ref{pair}) with the above identification.

\subsection{$S_3$-twisted Regular Punctures}\label{S3open}

\begin{figure}
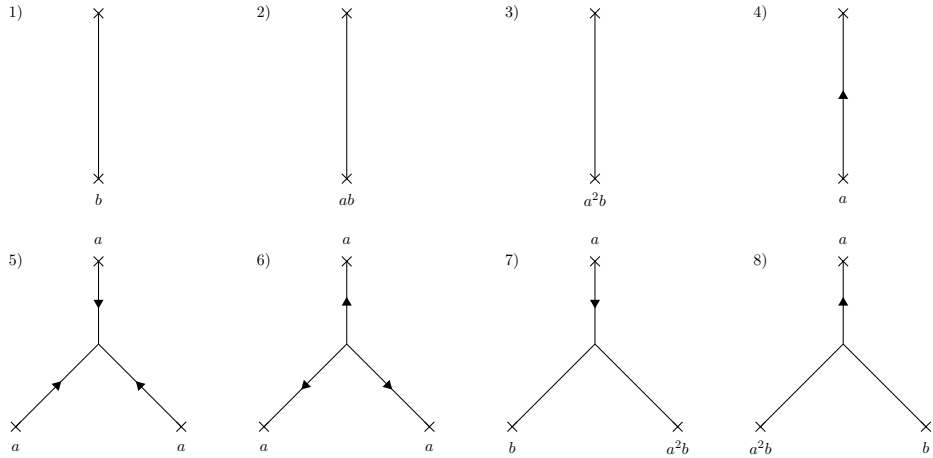

\centering
\scalebox{.55}{

}
\caption{Various kinds of irreducible configurations of open twist lines valued in $S_3$. We name these 1-8 as follows: open $b$ line, open $ab$ line, open $a^2b$ line, meson, baryon, anti-baryon, anti-mixed configuration, mixed configuration. These configurations are distinct except for the anti-mixed and mixed configuration, see figure \ref{fig:Redund5}.}
\label{fig:IrredConfig}
\end{figure}

Now we consider incorporating more general regular twisted punctures in the $D_4$ $(2,0)$ theory. We can have the following various irreducible configurations of twisted regular punctures as shown in figure \ref{fig:IrredConfig}:
\bit
\item Two punctures joined by a $\Z_2$ twist line implementing the transformation $b\in S_3$ as one crosses it. We refer to this configuration as the open $b$ line.
\item Two punctures joined by a $\Z_2$ twist line implementing the transformation $ab\in S_3$ as one crosses it. We refer to this configuration as the open $ab$ line.
\item Two punctures joined by a $\Z_2$ twist line implementing the transformation $a^2b\in S_3$ as one crosses it. We refer to this configuration as the open $a^2b$ line.
\item Two punctures joined by an oriented $\Z_3$ twist line implementing the transformation $a\in S_3$ as one crosses it in a particular direction (which is left to right in the fourth configuration of figure \ref{fig:IrredConfig}). We refer to this configuration as a ``meson''.
\item Three punctures acting as sources of three $a$ twist lines. The three twist lines meet at a point and annihilate each other. We refer to this configuration as a ``baryon''.
\item Three punctures acting as sinks of three $a$ twist lines. The three twist lines originate from a common point. We refer to this configuration as an ``anti-baryon''.
\item Two punctures emitting $a^2b$ and $b$ $\Z_2$ twist lines which combine to form an $a$ twist line which ends at a puncture. We refer to this configuration as a ``mixed'' configuration.
\item A puncture emitting an $a$ twist line which then splits into $a^2b$ and $b$ $\Z_2$ twist lines. Each $\Z_2$ twist line ends on a puncture. We refer to this configuration as an ``anti-mixed'' configuration.
\eit

\begin{figure}
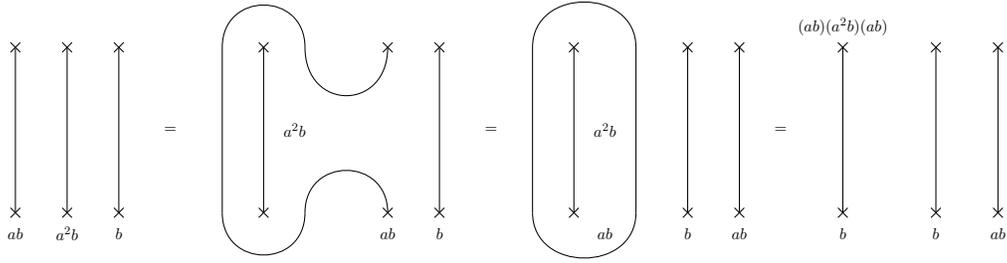

\centering
\scalebox{.55}{

}
\caption{A configuration involving three different types open $\Z_2$ twist lines can be topologically deformed to a configuration involving only two different types of open $\Z_2$ twist lines. Going from the third configuration to the last configuration involves fusing the closed $ab$ loop with the open $a^2b$ line, which conjugates $a^2b$ by $ab$, resulting in an open $b$ line.}
\label{fig:Redund1}
\end{figure}

\begin{figure}
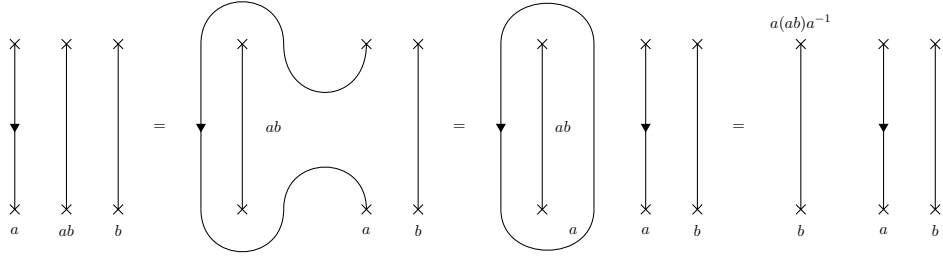

\centering
\scalebox{.55}{%
}
\caption{A configuration involving two different types open $\Z_2$ twist lines and a meson can be topologically deformed to a configuration involving a meson and open $\Z_2$ twist lines of a single type only. Going from the third configuration to the last configuration involves fusing the closed $a$ loop with the open $ab$ line, which conjugates $ab$ by $a$, resulting in an open $b$ line. }
\label{fig:Redund2}
\end{figure}

\begin{figure}
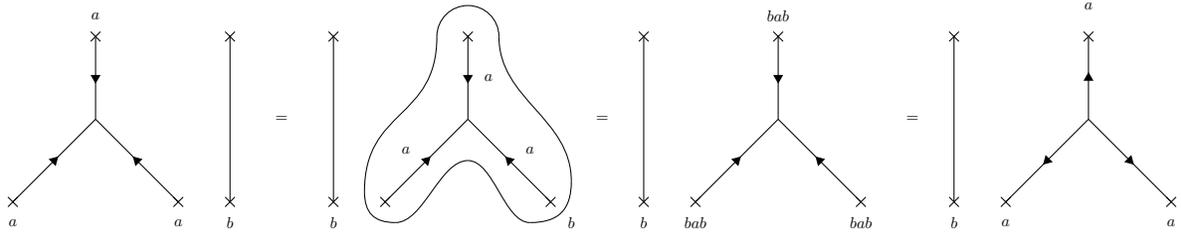

\centering
\scalebox{.55}{%
}
\caption{A baryon can be converted into an anti-baryon by passing it through a $b$ twist line.}
\label{fig:Redund3}
\end{figure}

\begin{figure}
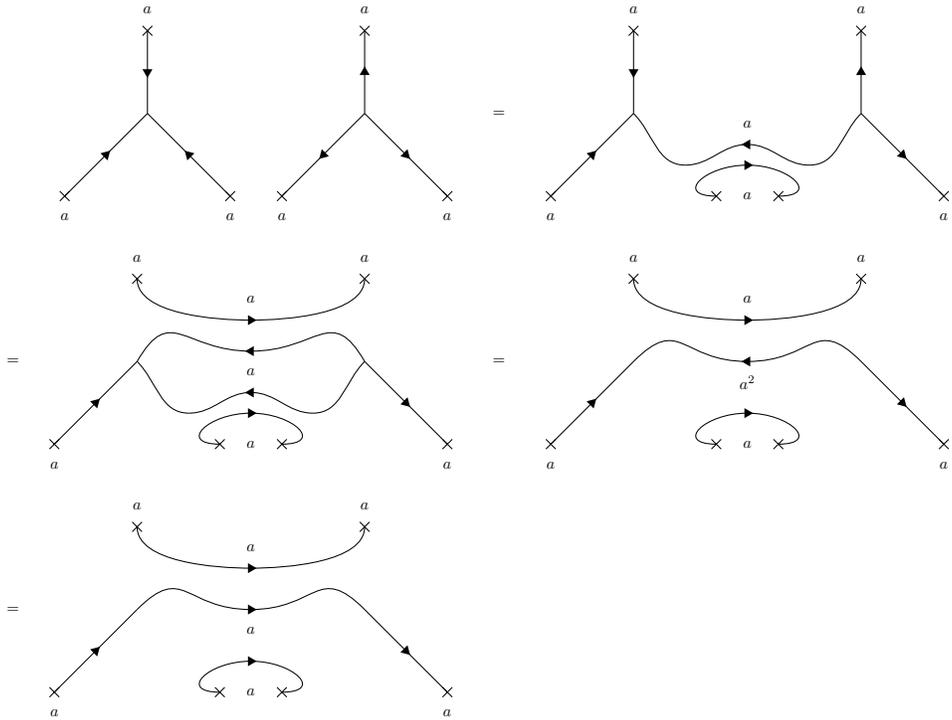

\centering
\scalebox{.55}{%
}
\caption{A configuration involving a baryon and an anti-baryon is topologically equivalent to a configuration involving three mesons.}
\label{fig:Redund4}
\end{figure}

\begin{figure}
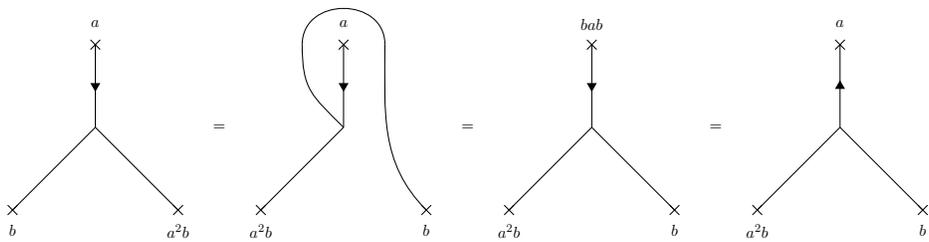

\centering
\scalebox{.55}{
%
}
\caption{An anti-mixed configuration is topologically equivalent to a mixed configuration.}
\label{fig:Redund5}
\end{figure}

\begin{figure}
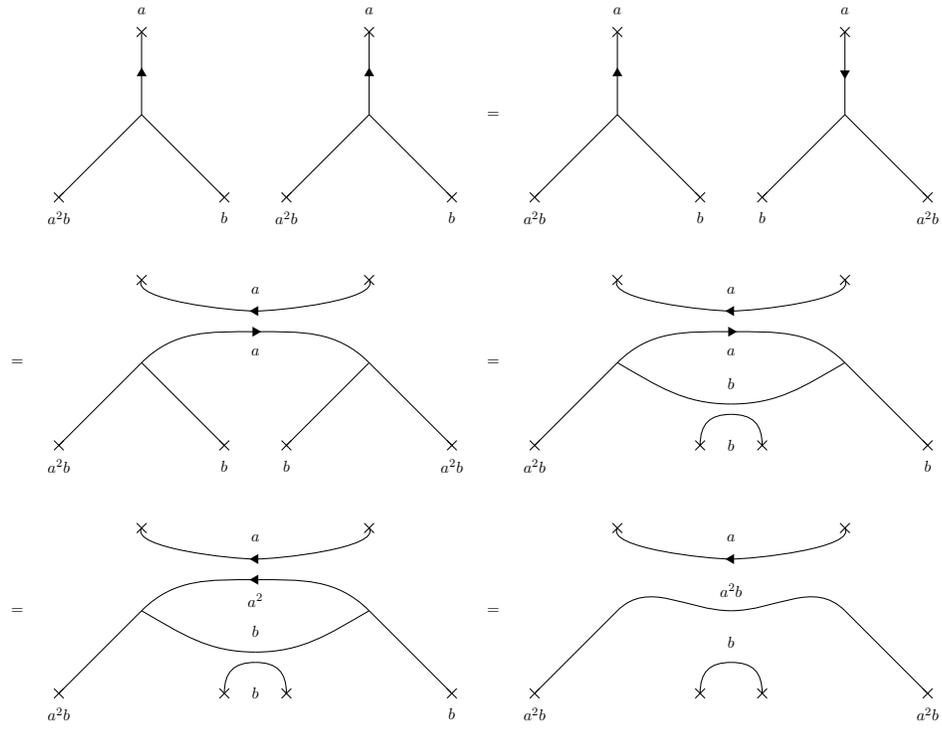

\centering
\scalebox{.55}{%
}
\caption{Two mixed configurations can be converted into a meson and two different kinds of open $\Z_2$ twist lines. Using figure \ref{fig:Redund2}, this is equivalent to a meson plus two open $b$ lines of the same type, since the last configuration can be further transformed according to figure \ref{fig:Redund2}.}
\label{fig:Redund6}
\end{figure}

\begin{figure}
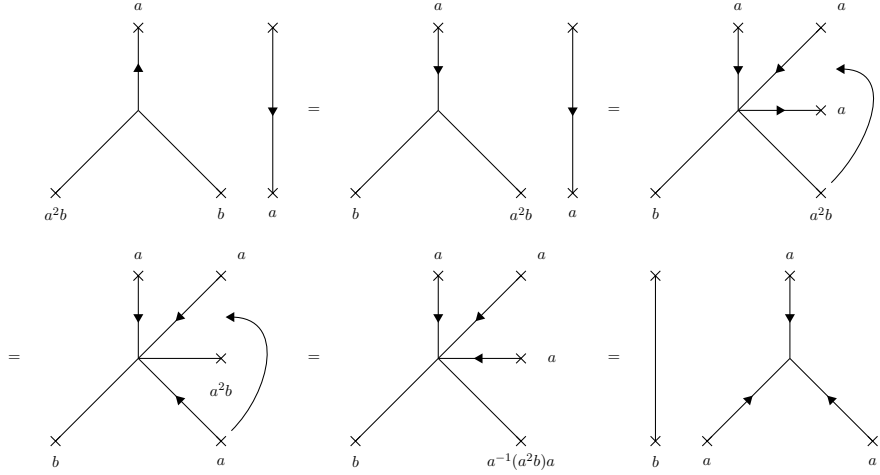

\centering
\scalebox{.55}{%
}
\caption{A mixed configuration plus a meson can be converted into a baryon plus an open $b$ line. We first convert the mixed configuration into an anti-mixed configuration. Then we fuse the open $a$ line internal to the meson with the junction for the mixed configuration to create a combined junction for all the open twist lines. Then we move a puncture living at the end of $a^2b$ line over a puncture acting as sink for $a$ line. This converts the latter puncture into a puncture acting as source for $a$ line (due to conjugation). Then we move this puncture over the puncture living at the end of $a^2b$ line, thus converting the latter into a puncture living at the end of a $b$ line. Finally we can separate a full open $b$ line from the junction leaving behind a baryon.}
\label{fig:Redund7}
\end{figure}

\begin{figure}
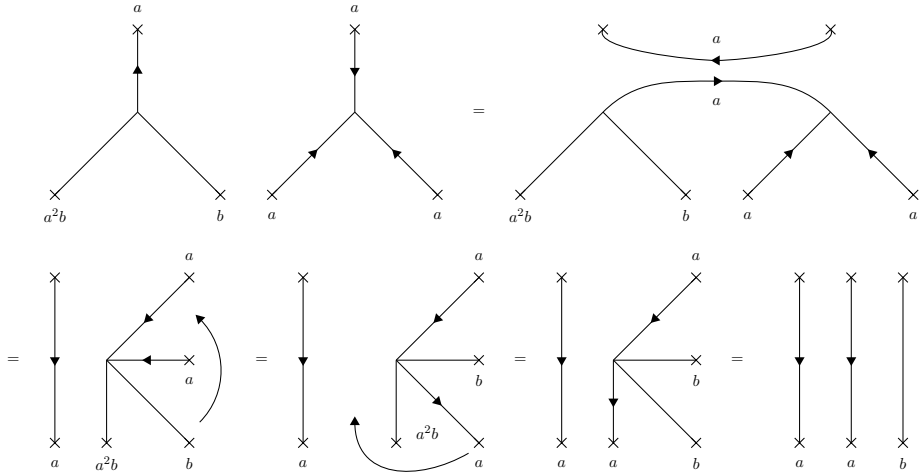

\centering
\scalebox{.55}{

}
\caption{A mixed configuration and a baryon can decomposed into two mesons and an open $b$ line.}
\label{fig:Redund8}
\end{figure}

\ni There are plenty of redundancies when we try to combine the above configurations:
\bit
\item Consider a situation where we have an open $b$ line, an open $ab$ line and an open $a^2b$ line. We can pass the open $a^2b$ line through the $ab$ line to convert the open $a^2b$ line into an open $b$ line. See figure \ref{fig:Redund1}. At the end of this process, we obtain a situation in which we have two open $b$ lines and one open $ab$ line.
\item Consider a situation where we have an open $a$ line, an open $b$ line and an open $ab$ line. We can pass the open $ab$ line through the $a$ line to convert the open $ab$ line into an open $b$ line. See figure \ref{fig:Redund2}. At the end of this process, we obtain a situation in which we have one open $a$ line and two open $b$ lines.
\item Consider a situation where we have a baryon and an open $b$ line. We can pass the baryon through the $b$ line to convert the baryon into an anti-baryon. See figure \ref{fig:Redund3}. At the end of this process, we obtain a situation in which we have an anti-baryon and an open $b$ line.
\item A baryon and an anti-baryon can be decomposed as three mesons. See figure \ref{fig:Redund4}.
\item An anti-mixed configuration can be converted into a mixed configuration. See figure \ref{fig:Redund5}.
\item Two mixed configurations can be decomposed as a meson and two open $b$ lines. See figure \ref{fig:Redund6}.
\item A mixed configuration plus a meson is equivalent to a baryon plus an open $b$ line. See figure \ref{fig:Redund7}.
\item A mixed configuration plus a baryon is equivalent to two mesons plus an open $b$ line. See figure \ref{fig:Redund8}.
\eit

\begin{figure}
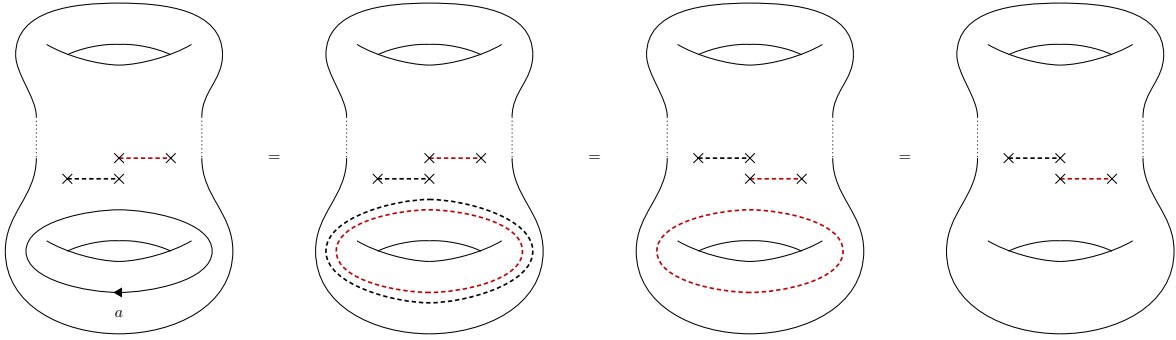

\centering
\scalebox{0.55}{
}
\caption{A configuration involving a closed $\Z_3$ twist line and two different types of open $\Z_2$ twist lines is topologically equivalent to a configuration involving only the two different types of open $\Z_2$ twist line without the closed $\Z_3$ twist line. The different types of $\Z_2$ twist lines are distinguished by different colors in the above figure.}
\label{fig:OpenZ3TwistLine}
\end{figure}

\ni Accounting for the above redundancies we can easily show that the only topologically distinct possibilities for non-trivial $S_3$ twist lines on $C_g$ are as follows:
\bit
\item $k$ open $b$ lines. This was discussed in the previous subsection.
\item $k$ open $b$ lines plus a $\Z_3$ closed twist line.
\item $k$ open $b$ lines plus $k'$ open $ab$ lines. Inserting an additional $\Z_3$ closed twist line does not lead to a topologically distinct scenario. See figure \ref{fig:OpenZ3TwistLine}.
\item $l$ mesons.
\item $l$ mesons plus a $\Z_2$ closed twist line.
\item $k$ open $b$ lines plus $l$ mesons.
\item $p$ baryons.
\item $p$ baryons plus $l$ mesons.
\item One baryon plus $l$ mesons plus a $\Z_2$ closed twist line.
\item One baryon plus $k$ open $b$ lines.
\item One baryon plus $k$ open $b$ lines plus $l$ mesons.
\item One mixed configuration.
\item One mixed configuration plus $k$ open $b$ lines.
\eit

\begin{figure}
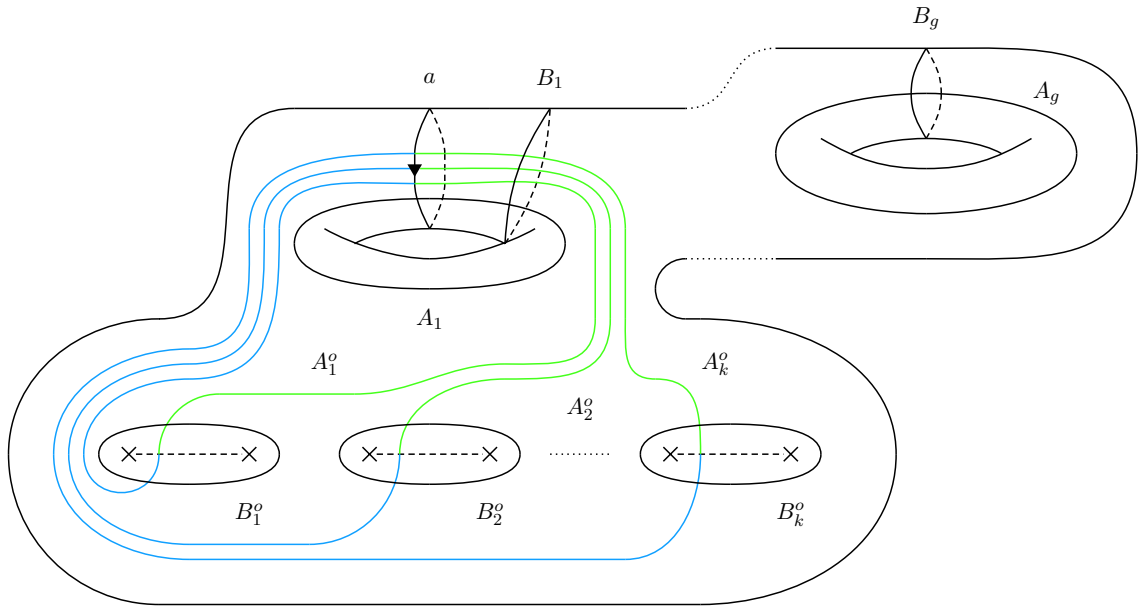

\centering
\scalebox{.8}{%
}
\caption{Riemann surface of genus $g$ with $k$ open $b$ lines and a closed $a$ line. Each cycle $A^i_o$ is broken into green and blue sub-segments lying between two difference kinds of twist lines the cycle crosses.}
\label{fig:TwistedRegularD4}
\end{figure}

We now determine $\cL$ for all of the above topologically distinct possibilities one-by-one. First consider the case where we have $k>0$ open $b$ lines and a closed $a$ line. We define cycles $A^o_i$ and $B^o_i$ as shown in figure \ref{fig:TwistedRegularD4}, and let the $4d$ line operators (modulo screening and flavor charges) originating from them be $\cL_i^A$ and $\cL_i^B$ respectively. As before, $\cL_i^B\simeq\Z_2$ which can be generated by wrapping $s$ along $B^o_i$. On the other hand, $\cL_i^A\simeq\Z_2$ as well, and the generator can be chosen to be $s$ wrapped along the green sub-segment and $c$ wrapped along the blue sub-segment. We call the generators of $\cL_i^A$ and $\cL_i^B$ as $g_i^A$ and $g_i^B$ respectively. Then we can write the set $\cL$ of $4d$ line operators (modulo screening and flavor charges) as
\be
\cL\simeq\prod_{i=1}^{k}\cL^A_{i}\times\prod_{i=1}^{k}\cL^B_i\times\wh{Z}^{g-1}_A\times\wh{Z}^{g-1}_B
\ee
with the non-trivial pairings being
\be
\langle g^A_{i},g^B_i\rangle=\half
\ee
along with the pairing on $\wh{Z}^{g-1}_A\times\wh{Z}^{g-1}_B$.

\begin{figure}
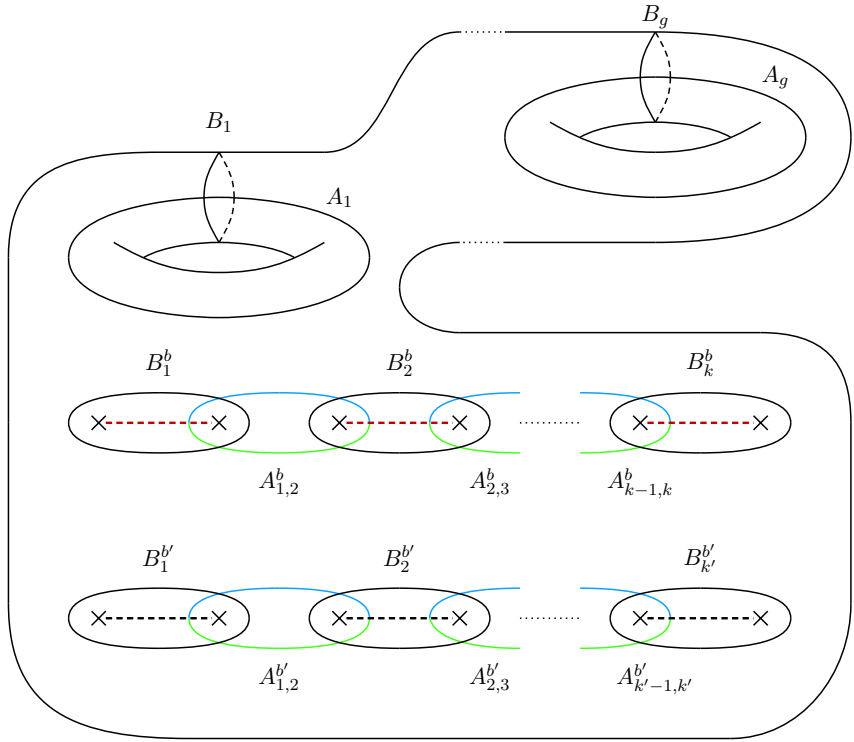

\centering
\scalebox{0.8}{
}
\caption{Riemann surface of genus $g$ with $k$ open $b$ lines and $k'$ open $b'$ lines with $b'\neq b$. The two different kinds of $\Z_2$ twisted open lines are displayed using different colors.}
\label{fig:TwoTwistLines}
\end{figure}

Now consider the case  where we have $k>0$ open $b$ lines and $k'>0$ open $b':=ab$ lines. See figure \ref{fig:TwoTwistLines}. Then the line operators arising from $B^{b'}_i$ can be generated by wrapping $v$ along $B^{b'}_i$. However, $v$ wrapped along $\sum_i B^{b'}_i$ is equivalent to $v$ wrapped along $\sum_i B^{b}_i$, which in turn can be trivialized as $v$ can be moved across twisted regular punctures of type $b$. Similarly, $s$ wrapped along $\sum_i B^{b}_i$ is trivial. Thus, we can write the set $\cL$ of $4d$ line operators (modulo screening and flavor charges) as
\be
\cL\simeq\prod_{i=1}^{k-1}\cL^{A,b}_{i,i+1}\times\prod_{i=1}^{k-1}\cL^{B,b}_i\times\prod_{i=1}^{k'-1}\cL^{A,b'}_{i,i+1}\times\prod_{i=1}^{k'-1}\cL^{B,b'}_i\times\wh{Z}^{g}_A\times\wh{Z}^{g}_B \,,
\ee
where $\cL^{A,b}_{i,i+1}\simeq\Z_2$ and $\cL^{A,b'}_{i,i+1}\simeq\Z_2$ are the sets of line operators descending from cycles $A^b_{i,i+1}$ and $A^{b'}_{i,i+1}$ respectively, and $\cL^{B,b}_i\simeq\Z_2$ and $\cL^{B,b'}_i\simeq\Z_2$ are the sets of line operators descending from cycles $B^b_{i}$ and $B^{b'}_{i}$ respectively. Let the corresponding generators be $g^{A,b}_{i,i+1}, g^{A,b'}_{i,i+1}, g^{B,b}_{i}, g^{B,b'}_{i}$. We can define $g^{A,b}_{i,i+1}$ by inserting $s$ on the green sub-segment of $A^b_{i,i+1}$ and $c$ on the blue sub-segment of $A^b_{i,i+1}$; $g^{A,b'}_{i,i+1}$ by inserting $c$ on the green sub-segment of $A^{b'}_{i,i+1}$ and $v$ on the blue sub-segment of $A^{b'}_{i,i+1}$. Then the non-trivial pairings are
\be\label{pairr}
\langle g^{A,b}_{i,i+1},g^{B,b}_i\rangle=\langle g^{A,b}_{i,i+1},g^{B,b}_{i+1}\rangle=\langle g^{A,b'}_{i,i+1},g^{B,b'}_i\rangle=\langle g^{A,b'}_{i,i+1},g^{B,b'}_{i+1}\rangle=\half
\ee
along with the pairing \eqref{eq:Pairing} on $\wh{Z}^{g}_A\times\wh{Z}^{g}_B$.

\begin{figure}
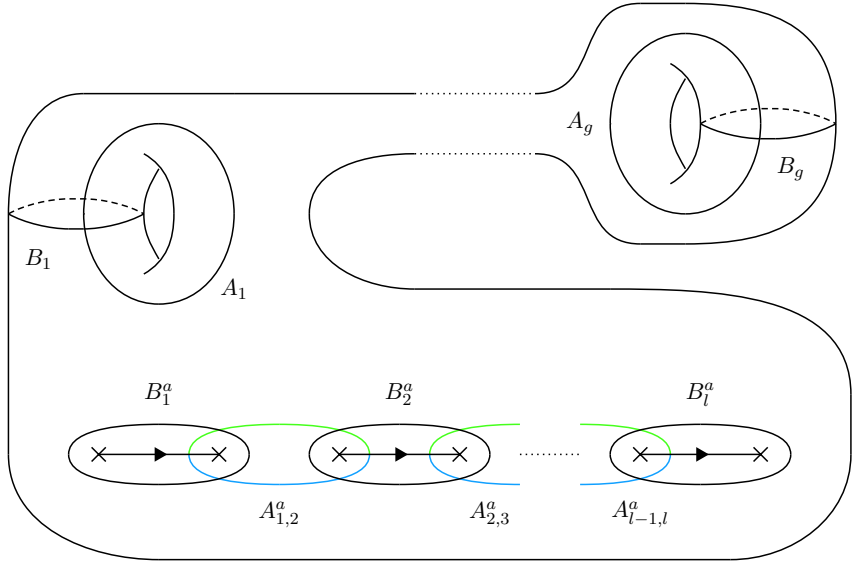

\centering
\scalebox{.8}{
}
\caption{Riemann surface of genus $g$ with $l$ mesons.}
\label{fig:SetUpMesonWithTwist2}
\end{figure}

Now consider the situation containing $l$ mesons only. See figure \ref{fig:SetUpMesonWithTwist2}. Then the line operators $\cL^{B,a}_i$ arising from the cycle $B^a_i$ can be identified with $\Z_2\times\Z_2$, since no element of $\wh{Z}$ can be moved across a $\Z_3$ twisted puncture. We label the line operator in $\cL^{B,a}_i$ arising by wrapping $s$ along $B^a_i$ as $s^{B,a}_i$, and the line operator in $\cL^{B,a}_i$ arising by wrapping $c$ along $B^a_i$ as $c^{B,a}_i$. We choose $s^{B,a}_i$ and $c^{B,a}_i$ as the generators for $\cL^{B,a}_i$. Finally, $\sum_i s^{B,a}_i = \sum_i c^{B,a}_i =0$. Similarly, the line operators $\cL^{A,a}_{i,i+1}$ arising from the cycle $A^a_{i,i+1}$ can be identified with $\Z_2\times\Z_2$. The element of $\cL^{A,a}_{i,i+1}$ arising by wrapping $s$ along the green sub-segment of $A^a_{i,i+1}$ is called $s^{A,a}_{i,i+1}$, and the element of $\cL^{A,a}_{i,i+1}$ arising by wrapping $v$ along the green sub-segment of $A^a_{i,i+1}$ is called $v^{A,a}_{i,i+1}$. We choose $s^{A,a}_{i,i+1}$ and $v^{A,a}_{i,i+1}$ as the generators for $\cL^{A,a}_{i,i+1}$. In total, we have
\be
\cL\simeq\prod_{i=1}^{l-1}\cL^{A,a}_{i,i+1}\times\prod_{i=1}^{l-1}\cL^{B,a}_i\times\wh{Z}^{g}_A\times\wh{Z}^{g}_B
\ee
with the non-trivial pairings being
\be\label{ntp}
\langle s^{A,a}_{i,i+1}, s^{B,a}_i\rangle=\langle s^{A,a}_{i,i+1}, s^{B,a}_{i+1}\rangle=\langle v^{A,a}_{i,i+1}, c^{B,a}_i\rangle=\langle v^{A,a}_{i,i+1}, c^{B,a}_{i+1}\rangle=\half
\ee
along with the pairing on $\wh{Z}^{g}_A\times\wh{Z}^{g}_B$.

\begin{figure}
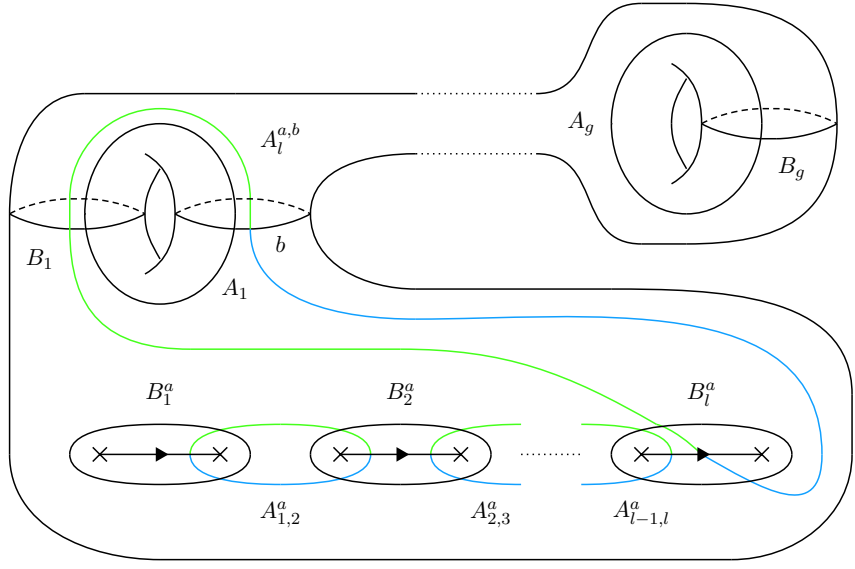

\centering
\scalebox{0.8}{

}
\caption{Riemann surface of genus $g$ with $l$ mesons and a closed $b$ line.}
\label{fig:MesonWithTwist2}
\end{figure}

Now let us consider $l$ mesons in the presence of a closed $\Z_2$ twist line of type $b$. See figure \ref{fig:MesonWithTwist2}. Notice that now $\sum_i s^{B,a}_i = \sum_i c^{B,a}_i\neq0$. Thus, we label the $\Z_2$ subgroup of $\cL^{B,a}_l$ generated by $s^{B,a}_l$ as $\cS^{B,a}_l$. Correspondingly, there is a new cycle $A^{a,b}_l$ shown in figure \ref{fig:MesonWithTwist2} giving rise to a $4d$ line operator, which is obtained by wrapping $c$ along the green sub-segment and $s$ along the blue sub-segment of $A^{a,b}_l$. We label this set of line operators as $\cL^{A,a,b}_l\simeq\Z_2$ and its generator described above as $c^{A,a,b}_l$. We choose the generator of the set of line operators $\text{Proj}(\wh{Z},o)\simeq\Z_2$ originating from cycle $B_1$ shown in figure \ref{fig:MesonWithTwist2} to be $c$ wrapping the cycle $B_1$. Then, we obtain that the total set of $4d$ line operators is
\be
\cL\simeq \prod_{i=1}^{l-1}\cL^{A,a}_{i,i+1}\times\prod_{i=1}^{l-1}\cL^{B,a}_i\times\cL^{A,a,b}_l\times\cS^{B,a}_l\times\text{Inv}(\wh{Z}, o) \times \text{Proj}(\wh{Z},o)\times\wh{Z}^{g-1}_A\times\wh{Z}^{g-1}_B
\ee
with non-trivial pairings (\ref{ntp}), along with the pairing on $\wh{Z}^{g-1}_A\times\wh{Z}^{g-1}_B$ and the new pairings
\be
\langle s^{A,a}_{l-1,l}, s^{B,a}_l\rangle=\langle c^{A,a,b}_{l}, s^{B,a}_{l}\rangle=\half \,.
\ee

Consider now $l$ mesons with $2k\neq0$ $\Z_2$ twisted regular punctures of type $b$. We have the constraint that $\sum_i s^{B,a}_i = \sum_i c^{B,a}_i = \sum_i g^{B,b}_i$. Also we have a new cycle $A^{b,a}_{k,1}$ as shown in figure \ref{fig:MesonWithTwistPuncture2} which contributes a group $\cL^{A,b,a}_{k,1}\simeq\Z_2$ of $4d$ line operators which is generated by $g^{A,b,a}_{k,1}$ which is obtained by wrapping $c$ along the green sub-segment and $s$ along the blue sub-segment of $A^{b,a}_{k,1}$. In total, we have
\be
\cL\simeq\prod_{i=1}^{k-1}\cL^{A,b}_{i,i+1}\times \cL^{A,b,a}_{k,1}\times\prod_{i=1}^{k}\cL^{B,b}_i\times\prod_{i=1}^{l-1}\cL^{A,a}_{i,i+1}\times\prod_{i=1}^{l-1}\cL^{B,a}_i\times\wh{Z}^{g}_A\times\wh{Z}^{g}_B
\ee
The non-trivial pairings are those on $\wh{Z}^{g}_A\times\wh{Z}^{g}_B$, those given in (\ref{ntp}), and those listed below
\be
\langle g^{A,b}_{i,i+1},g^{B,b}_i\rangle=\langle g^{A,b}_{i,i+1},g^{B,b}_{i+1}\rangle=\langle g^{A,b,a}_{k,1},g^{B,b}_k\rangle=\langle g^{A,b,a}_{k,1},s^{B,a}_1\rangle=\langle g^{A,b,a}_{k,1},c^{B,a}_1\rangle=\half \,.
\ee

\begin{figure}
\centering
\scalebox{.55}{
}
\caption{Riemann surface of genus $g$ with $k$ open $b$ lines and $l$ mesons.}
\label{fig:MesonWithTwistPuncture2}
\end{figure}

\begin{figure}
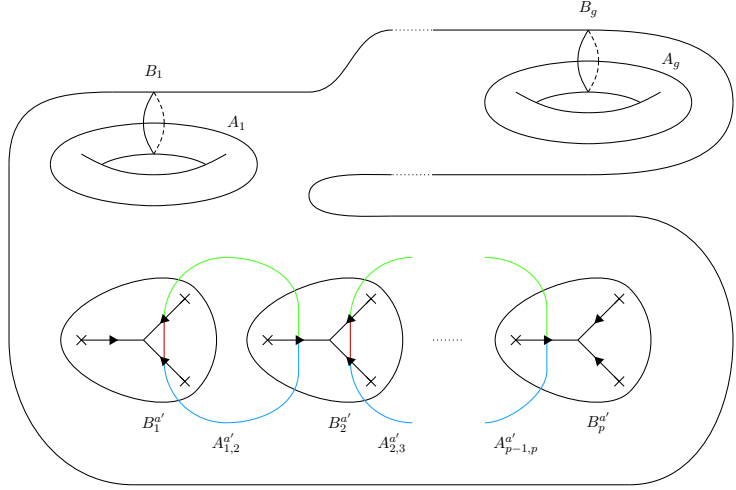

\centering
\scalebox{.55}{%
}
\caption{Riemann surface of genus $g$ with $p$ baryons.}
\label{fig:PBaryons}
\end{figure}

Consider now the case involving $p$ baryons. See figure \ref{fig:PBaryons}. Let $\cL^{A,a'}_{i,i+1}$ be the set of $4d$ line operators arising from the cycle $A^{a'}_{i,i+1}$ and $\cL^{B,a'}_{i}$ be the set of $4d$ line operators arising from the cycle $B^{a'}_{i}$. We can wrap any element of $\wh{Z}\simeq\Z_2\times\Z_2$ along $B^{a'}_{i}$ which implies that $\cL^{B,a'}_{i}\simeq\Z_2\times\Z_2$. We choose the generators of $\cL^{B,a'}_i$ to be $s^{B,a'}_i$ and $c^{B,a'}_i$, which are obtained by wrapping $s$ and $c$ respectively along $B^{a'}_{i}$. Similarly, $\cL^{A,a'}_{i,i+1}\simeq\Z_2\times\Z_2$. The element of $\cL^{A,a'}_{i,i+1}$ arising by wrapping $s$ along the green sub-segment of $A^{a'}_{i,i+1}$ (which implies that $v$ is wrapped along the red sub-segment and $c$ is wrapped along the blue sub-segment) is called $s^{A,a'}_{i,i+1}$, and the element of $\cL^{A,a'}_{i,i+1}$ arising by wrapping $v$ along the green sub-segment of $A^{a'}_{i,i+1}$ is called $v^{A,a'}_{i,i+1}$. We choose $s^{A,a'}_{i,i+1}$ and $v^{A,a'}_{i,i+1}$ as the generators for $\cL^{A,a'}_{i,i+1}$. In total, we have
\be
\cL\simeq\prod_{i=1}^{p-1}\cL^{A,a'}_{i,i+1}\times\prod_{i=1}^{p-1}\cL^{B,a'}_i\times\wh{Z}^{g}_A\times\wh{Z}^{g}_B
\ee
with the non-trivial pairings being
\be
\langle s^{A,a'}_{i,i+1}, s^{B,a'}_i\rangle=\langle s^{A,a'}_{i,i+1}, s^{B,a'}_{i+1}\rangle=\langle v^{A,a'}_{i,i+1}, c^{B,a'}_i\rangle=\langle v^{A,a'}_{i,i+1}, c^{B,a'}_{i+1}\rangle=\half
\ee
along with the pairing on $\wh{Z}^{g}_A\times\wh{Z}^{g}_B$.

\begin{figure}
\centering
\scalebox{.55}{

}
\caption{Riemann surface of genus $g$ with $l$ mesons and $p$ baryons.}
\label{fig:PBaryonsLMesons}
\end{figure}

Consider now the case involving $p$ baryons and $l$ mesons. See figure \ref{fig:PBaryonsLMesons}. Along with the previously discussed groups $\cL^{A,a}_{i,i+1},\cL^{A,a'}_{i,i+1},\cL^{B,a}_{i},\cL^{B,a'}_{i}$, we also have a group $\cL^{A,a,a'}_{l,1}$ arising from the cycle $A^{a,a'}_{l,1}$ shown in figure \ref{fig:PBaryonsLMesons}. We have $\cL^{A,a,a'}_{l,1}\simeq\Z_2\times\Z_2$. The element of $\cL^{A,a,a'}_{l,1}$ arising by wrapping $s$ along the green sub-segment of $A^{a,a'}_{l,1}$ is called $s^{A,a,a'}_{l,1}$, and the element of $\cL^{A,a,a'}_{l,1}$ arising by wrapping $v$ along the green sub-segment of $A^{a,a'}_{l,1}$ is called $v^{A,a,a'}_{l,1}$. We choose $s^{A,a,a'}_{l,1}$ and $v^{A,a,a'}_{l,1}$ as the generators for $\cL^{A,a,a'}_{l,1}$. In total, we have
\be
\cL\simeq\prod_{i=1}^{l-1}\cL^{A,a}_{i,i+1}\times\cL^{A,a,a'}_{l,1}\times\prod_{i=1}^{l}\cL^{B,a}_i\times\prod_{i=1}^{p-1}\cL^{A,a'}_{i,i+1}\times\prod_{i=1}^{p-1}\cL^{B,a'}_i\times\wh{Z}^{g}_A\times\wh{Z}^{g}_B
\ee
with
\be
\langle s^{A,a,a'}_{l,1}, s^{B,a}_l\rangle=\langle s^{A,a,a'}_{l,1}, s^{B,a'}_{1}\rangle=\langle v^{A,a,a'}_{l,1}, c^{B,a}_l\rangle=\langle v^{A,a,a'}_{l,1}, c^{B,a'}_{1}\rangle=\half
\ee
being the new non-trivial pairings.

\begin{figure}
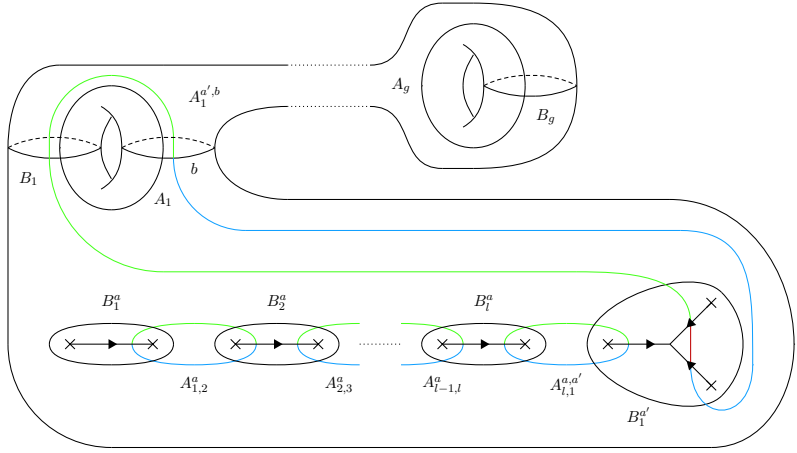

\centering
\scalebox{.55}{

}
\caption{Riemann surface of genus $g$ with $l$ mesons, one baryon and a closed $b$ line.}
\label{fig:PBaryonsLMesonsClosedb}
\end{figure}

We now consider the case having a single baryon, $l$ mesons and a closed $b$ line. See figure \ref{fig:PBaryonsLMesonsClosedb}. The only cycle not discussed above is $A_1^{a',b}$ which gives rise to a set of $4d$ line operators $\cL^{A,a',b}_1\simeq\Z_2$ whose generator $c_1^{A,a',b}$ is obtained by wrapping $c$ along the green sub-segment of $A_1^{a',b}$. Moreover, we can write $c_1^{B,a'}=\sum_{i=1}^l s_i^{B,a}+s_1^{B,a'}+\sum_{i=1}^l c_i^{B,a}$ implying that the relevant set of $4d$ line operators arising from $B_1^{a'}$ can be taken to be $\cS^{B,a'}_1\simeq\Z_2$ which is generated by $s^{B,a'}_1$. In total, we have
\be
\cL\simeq\prod_{i=1}^{l-1}\cL^{A,a}_{i,i+1}\times\cL^{A,a,a'}_{l,1}\times\prod_{i=1}^{l}\cL^{B,a}_i\times\cL^{A,a',b}\times\cS^{B,a'}_1\times\wh{Z}^{g-1}_A\times\wh{Z}^{g-1}_B
\ee
with
\be
\langle c_1^{A,a',b}, s^{B,a'}_1\rangle=\half
\ee
being the only new non-trivial pairing not discussed previously.

\begin{figure}
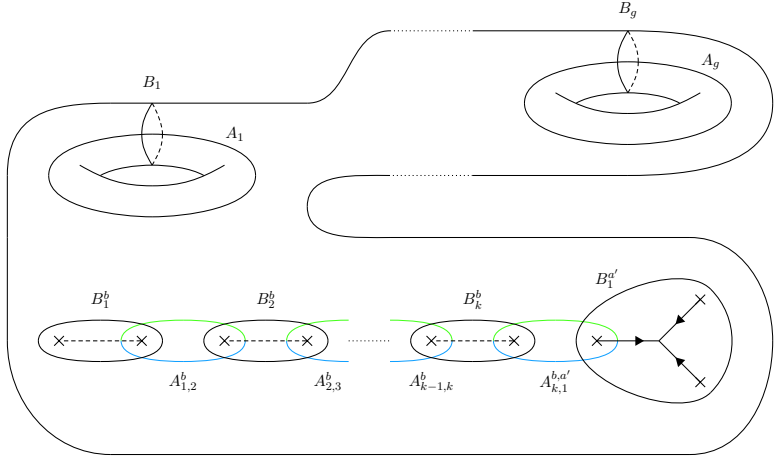

\centering
\scalebox{.55}{

}
\caption{Riemann surface of genus $g$ with $k$ open $b$ lines and one baryon.}
\label{fig:BaryonKOpenLines}
\end{figure}

We now consider the case having a single baryon and $k$ open $b$ lines. See figure \ref{fig:BaryonKOpenLines}. The only cycle not discussed above is $A_{k,1}^{b,a'}$ which gives rise to a set of $4d$ line operators $\cL^{A,b,a'}_{k,1}\simeq\Z_2$ whose generator $g_{k,1}^{A,b,a'}$ is obtained by wrapping $c$ along the green sub-segment of $A_{k,1}^{b,a'}$. Moreover, we can write $c_1^{B,a'}=s_1^{B,a'}=\sum_{i=1}^k g^{B,b}_i$. In total, we have
\be
\cL\simeq\prod_{i=1}^{k-1}\cL^{A,b}_{i,i+1}\times\cL^{A,b,a'}_{k,1}\times\prod_{i=1}^{k}\cL^{B,b}_i\times\wh{Z}^{g}_A\times\wh{Z}^{g}_B
\ee
with
\be
\langle g_{k,1}^{A,b,a'},g_{k}^{B,b}\rangle=\half
\ee
being the only new non-trivial pairing not discussed previously.

\begin{figure}
\centering
\scalebox{.65}{

}
\caption{Riemann surface of genus $g$ with $k$ open $b$ lines, $l$ mesons and one baryon.}
\label{fig:BaryonsKLinesLmesons}
\end{figure}

Consider now the case involving a single baryon, $k$ open $b$ lines and $l$ mesons. See figure \ref{fig:BaryonsKLinesLmesons}. We can quickly deduce that
\be
\cL\simeq\prod_{i=1}^{k-1}\cL^{A,b}_{i,i+1}\times\cL^{A,b,a}_{k,1}\times\prod_{i=1}^{k}\cL^{B,b}_i\times\prod_{i=1}^{l-1}\cL^{A,a}_{i,i+1}\times\cL^{A,a,a'}_{l,1}\times\prod_{i=1}^{l}\cL^{B,a}_i\times\wh{Z}^{g}_A\times\wh{Z}^{g}_B
\ee
There are no new non-trivial pairings.

\begin{figure}
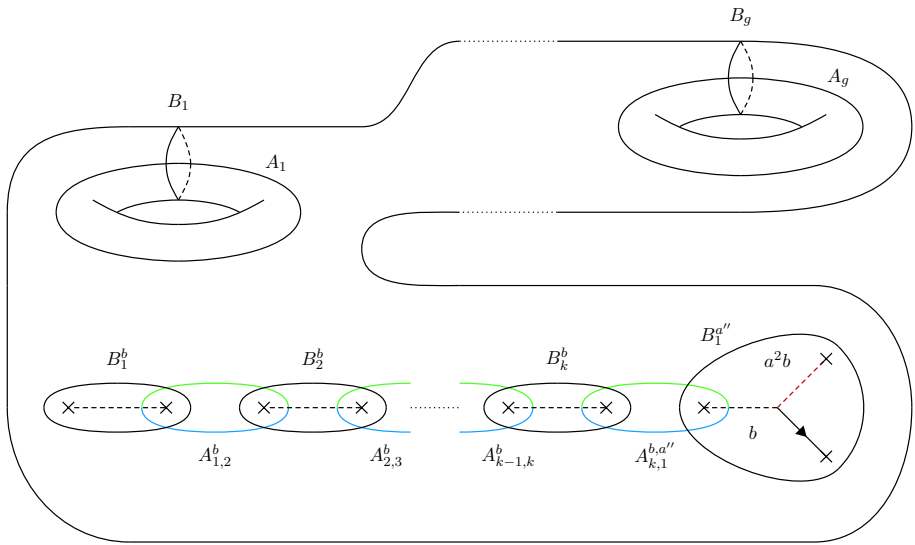

\centering
\scalebox{.65}{

}
\caption{Riemann surface of genus $g$ with $k$ open $b$ lines and one mixed configuration.}
\label{fig:MixedConfigKLines}
\end{figure}

Consider now the final case involving a single mixed configuration along with $k\ge0$ open $b$ lines. See figure \ref{fig:MixedConfigKLines}. We have $s(B_1^{a''})=c(B_1^{a''})=\sum_{i=1}^k g_i^{B,b}$. Thus $B_1^{a''}$ contributes no new $4d$ line operators. Moreover, we can obtain a non-trivial $4d$ line operator $g^{A,b,a''}_{k,1}$ by wrapping $c$ along the green sub-segment of $A^{b,a''}_{k,1}$. In total, we have
\be
\cL\simeq\prod_{i=1}^{k-1}\cL^{A,b}_{i,i+1}\times\cL^{A,b,a''}_{k,1}\times\prod_{i=1}^{k}\cL^{B,b}_i\times\wh{Z}^{g}_A\times\wh{Z}^{g}_B
\ee
with
\be
\langle g_{k,1}^{A,b,a''},g_{k}^{B,b}\rangle=\half
\ee
being the only new non-trivial pairing not discussed previously. $\cL$ is trivial for $k=0$.

We finish this subsection by discussing some Lagrangian examples. Some more examples illustrating the results of this subsection appear in the next subsection on atypical punctures.

\vspace{6mm}

\ni\ubf{Sphere with 2 $\Z_2$ twisted regular punctures of type $b$ and 2 $\Z_2$ twisted}\\
\ubf{regular punctures of type $b'\neq b$}: The $4d$ $\cN=2$ theory $\fg_2+4\F$ (carrying $\fg_2$ gauge algebra and 4 full hypers in irrep of dimension $\mathbf{7}$) can be constructed using a compactification of $D_4$ $(2,0)$ theory on a sphere with 4 regular twisted punctures, 2 open $\Z_2$ twist lines of type $b$ and 1 open $\Z_3$ twist line of type $a$ as shown in figure \ref{fig:G2F4GaiottoCurve} \cite{Tachikawa:2010vg}. Combining the two open $\Z_2$ twist lines, we obtain a configuration with 4 regular twisted punctures, 1 open $\Z_2$ twist line of type $b$ and 1 open $\Z_2$ twist line of type $b'\neq b$. See figure \ref{fig:G2F4GaiottoCurve}. Using our results above, we would thus conclude that $\cL$ should be trivial. Indeed, this is easily verified from the Lagrangian $\fg_2+4\F$ description.

\begin{figure}
\centering
\scalebox{.7}{

}
\caption{Converting a configuration of open twist lines on a sphere discussed in \cite{Tachikawa:2010vg} to a configuration of open twist lines discussed in this paper.}
\label{fig:G2F4GaiottoCurve}
\end{figure}

The $4d$ $\cN=2$ gauge theories $\so(8)+3\F+3\S$ and $\so(7)+2\F+3\S$ can be obtained by compactifying $D_4$ $(2,0)$ theory on a sphere with 4 regular twisted punctures, 2 open $\Z_2$ twist lines of type $b$ and 1 \emph{closed} $\Z_3$ twist line of type $a$ as shown in figure \ref{fig:SO8GaiottoCurve} \cite{Tachikawa:2010vg}. We can move the closed $\Z_3$ twist line such that it encircles 2 regular $\Z_2$ twisted punctures as shown in figure \ref{fig:SO8GaiottoCurve}. This corresponds to conjugating the enclosed open $\Z_2$ twist line $b$ by $a$. Thus, we can remove the closed $a$ twist line if we convert this open $\Z_2$ twist line from type $b$ to type $b'$. See figure \ref{fig:SO8GaiottoCurve}. This is the same configuration that we obtained above, and hence we expect that $\cL$ should be trivial. Indeed, this is the case for the two $4d$ $\cN=2$ gauge theories $\so(8)+3\F+3\S$ and $\so(7)+2\F+3\S$.

\begin{figure}
\centering
\scalebox{.7}{

}
\caption{Converting a configuration of open twist lines on a sphere discussed in \cite{Tachikawa:2010vg} to a configuration of open twist lines discussed in this paper.}
\label{fig:SO8GaiottoCurve}
\end{figure}

\subsection{Atypical Regular Punctures}\label{AP}
Atypical regular punctures can be straightforwardly included in our analysis by resolving each atypical regular puncture into typical regular punctures. See the beginning of Section \ref{w/p} for the definition of atypical regular punctures and further references which discuss them in detail.

\begin{figure}
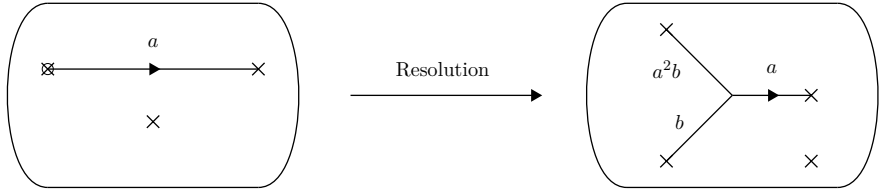

\centering
\scalebox{.7}{

}
\caption{Left: Compactification on a sphere involving a typical untwisted regular puncture, a typical twisted regular puncture and an \emph{atypical} twisted regular puncture. The atypical punctures are denoted by a circle super-imposed on top of a cross, while typical punctures are denoted by a cross only. Right: The atypical puncture is resolved to two typical twisted regular punctures. The resolution results in a mixed configuration.}
\label{fig:Resolution1}
\end{figure}

\vspace{6mm}

\ni\ubf{Gauge theory fixtures of type $(1,\omega,\omega^2)$}: We can obtain the $4d$ $\cN=2$ gauge theory $\sp(2)+6\F$ by compactifying $D_4$ $(2,0)$ theory on a sphere with one typical untwisted regular puncture, one typical twisted regular puncture acting as the sink of an $a$ twist line, and one \emph{atypical} twisted regular puncture acting as the source of an $a$ twist line \cite{Chacaltana:2016shw}, as shown in figure \ref{fig:Resolution1}. The atypical puncture can be resolved into two $\Z_2$ twisted typical regular punctures.  After this resolution, we observe that we have a sphere with what we referred to as a ``mixed'' configuration in section \ref{S3open}. Our analysis there suggests that we should have a trivial $\cL$, which matches the result obtained using the $\sp(2)+6\F$ gauge theory description.

\begin{figure}
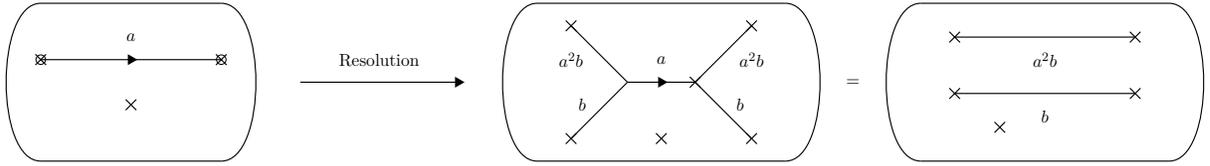

\centering
\scalebox{.6}{

}
\caption{Left: Compactification on a sphere involving a typical untwisted regular puncture and two \emph{atypical} twisted regular punctures. Center: Each atypical puncture is individually resolved into two typical twisted regular punctures. Right: After a topological manipulation, we find two open $\Z_2$ twist lines of different types.}
\label{fig:Resolution2}
\end{figure}

As another example, we can obtain the $4d$ $\cN=2$ gauge theory 
\be
%
\ee
by compactifying $D_4$ $(2,0)$ theory on a sphere with one typical untwisted regular puncture, one \emph{atypical} twisted regular puncture acting as the sink of an $a$ twist line, and one \emph{atypical} twisted regular puncture acting as the source of an $a$ twist line \cite{Chacaltana:2016shw}. See figure \ref{fig:Resolution2}. The atypical puncture acting as the source can be resolved into two $\Z_2$ twisted typical regular punctures, and the atypical puncture acting as the sink can be resolved into two $\Z_2$ twisted typical regular punctures plus one untwisted typical regular puncture. See figure \ref{fig:Resolution2}. After this resolution, we observe that we have a sphere containing two different kinds of open $\Z_2$ twist lines: an open $b$ line and an open $b':=a^2b$ line. Thus, from our analysis in section \ref{S3open} we expect to obtain a trivial $\cL$, which matches the result obtained using the above gauge theory description, as the reader can readily verify.

\begin{figure}
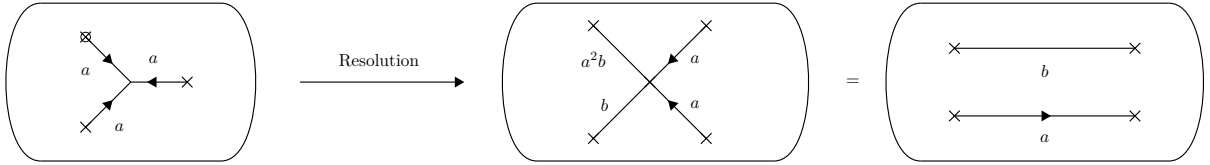

\centering
\scalebox{.6}{

}
\caption{Left: Compactification on a sphere involving two typical twisted regular punctures and one \emph{atypical} twisted regular puncture. Center: The atypical puncture is resolved into two typical twisted regular punctures. Right: After a topological manipulation, described in Figure \ref{fig:Redund8}, we end up with an open $b$ line and a meson.}
\label{fig:Resolution3}
\end{figure}

\vspace{6mm}

\ni\ubf{Gauge theory fixtures of type $(\omega,\omega,\omega)$}: We can obtain the $4d$ $\cN=2$ gauge theory $\sp(3)+2\L^2$ by compactifying $D_4$ $(2,0)$ theory on a sphere with three twisted regular punctures acting as sources of $a$ twist lines, thus forming a ``baryon-like'' configuration \cite{Chacaltana:2016shw}. See figure \ref{fig:Resolution3}. Two out of these three punctures are typical, while one of them is atypical. The atypical puncture can be resolved into two $\Z_2$ twisted typical regular punctures. See figure \ref{fig:Resolution3}. After this resolution, we observe that we have a sphere containing a configuration that we dealt with in figure \ref{fig:Redund8}. From the result of that figure, we know that this is equivalent to a sphere containing a meson-like configuration and an open $b$ line. Thus, from our analysis in section \ref{S3open} we expect to obtain
\be
\cL\simeq\Z_2\times\Z_2\,,
\ee
which matches the result obtained using the above gauge theory description, as the reader can readily verify.

\begin{figure}
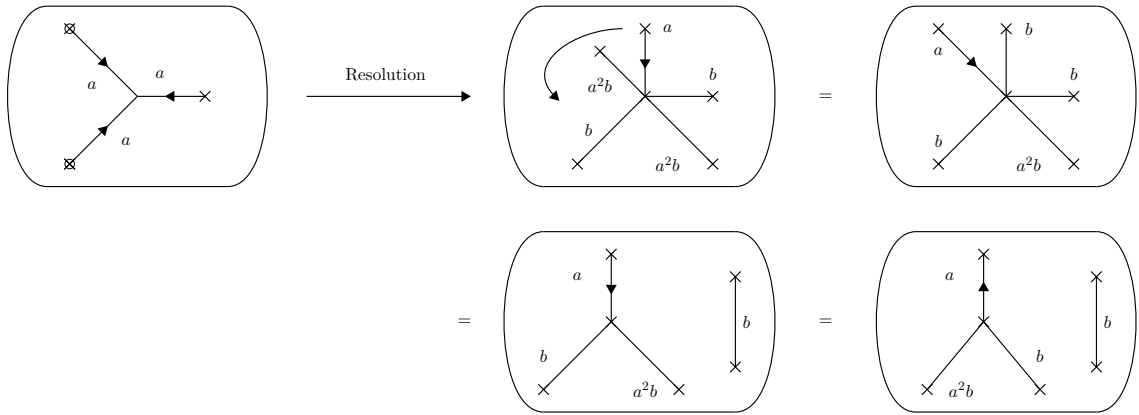

\centering
\scalebox{.6}{

}
\caption{Before the arrow: Compactification on a sphere involving one typical twisted regular puncture and two \emph{atypical} twisted regular punctures. After the arrow: Each atypical puncture is resolved into two typical twisted regular punctures. After some topological manipulations, we end up with an open $b$ line and a mixed configuration.}
\label{fig:Resolution4}
\end{figure}

As another example, we can obtain the $4d$ $\cN=2$ gauge theory with gauge algebra $\sp(2)\oplus\sp(2)$ and hypermultiplet content $\half(\L^2,\F)+(\L^2,1)+\frac72(1,\F)$ by compactifying $D_4$ $(2,0)$ theory on a sphere with three twisted regular punctures acting as sources of $a$ twist lines, thus forming a ``baryon-like'' configuration \cite{Chacaltana:2016shw}. See figure \ref{fig:Resolution4}. One out of these three punctures is typical, while two of them are atypical. Each atypical puncture can be resolved into two $\Z_2$ twisted typical regular punctures. See figure \ref{fig:Resolution4}. After this resolution, we can perform some topological moves, as shown in figure \ref{fig:Resolution4}, and reduce to a mixed configuration plus an open $b$ line. Thus, from our analysis in section \ref{S3open} we expect to obtain
\be
\cL\simeq\Z_2\times\Z_2\,,
\ee
which matches the result obtained using the above gauge theory description, as the reader can readily verify.

\section{Towards Irregular Punctures}\label{irr}

The analysis of this paper has focused on compactifications of $6d$ $(2,0)$ theories involving only regular (either untwisted or twisted) punctures. In this section, we discuss how our analysis can be generalized to incorporate irregular punctures, which are the punctures where the Hitchin field has poles of higher-order than simple poles.

A class of irregular punctures for $A_{n-1}$ $(2,0)$ theory were discussed in \cite{Nanopoulos:2010zb}. The Hitchin field at an irregular puncture of type $\cP_k$ in this class can be written as
\begin{align}
\varphi=&\frac1{z^{1+\frac1{n-k}}}\text{diag}(0,\cdots,0,\Lambda,\Lambda\omega,\cdots,\Lambda\omega^{n-k-1})dz\nn\\
&+\frac1z\text{diag}(m_1,m_2,\cdots,m_k,m_{k+1},m_{k+1},\cdots,m_{k+1})dz+\cdots
\end{align}
where $0\le k\le n-2$ and the mass parameters satisfy $\sum_{i=1}^{k+1} m_i=0$. Here $\omega$ is an $n$-th root of unity and $\Lambda$ denotes the dynamically generated scale. For irregular puncture of type $\cP_{n-1}$, we can instead write
\be
\varphi=\frac1{z^2}\text{diag}(\Lambda,\Lambda,\cdots,\Lambda,-(n-1)\Lambda)dz+\frac1z\text{diag}(m_1,m_2,\cdots,m_n)dz+\cdots
\ee
where $\sum_{i=1}^{n} m_i=0$.

\begin{figure}
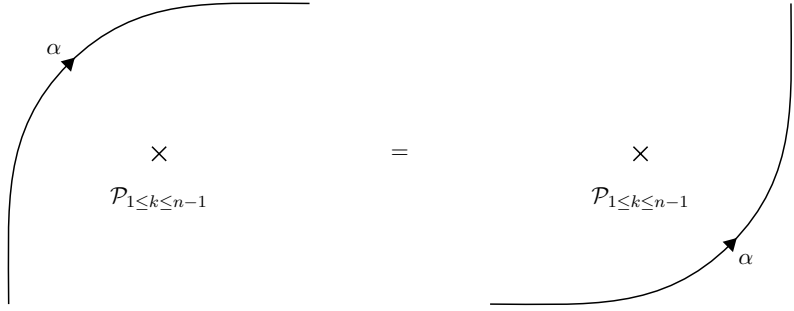

\centering
\scalebox{.80}{

}
\caption{A line carrying $\alpha\in\wh{Z}$ can be deformed across an irregular puncture of type $\cP_k$ for $1\le k\le n-1$.}
\label{fig:PkMove}
\end{figure}

\begin{figure}
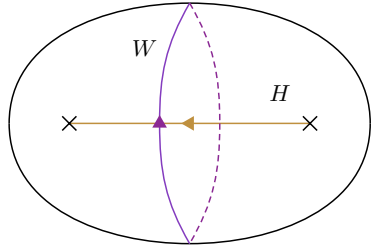

\centering
\scalebox{.80}{
%
}
\caption{A sphere two irregular punctures, both of type $\cP_k$. Two cycles $W$ and $H$ on this punctured sphere have been displayed.}
\label{fig:WH}
\end{figure}

We would now like to understand how these irregular punctures impact the determination of 1-form symmetry. In particular, we would like to understand whether an element $\alpha$ of $\wh{Z}\simeq\Z_n$ can be moved across an irregular puncture of type $\cP_k$ (where $k$ can take values in $\{0,1,\cdots,n-1\}$). To answer this question, we consider compactifying $A_{n-1}$ $(2,0)$ theory on a sphere with two irregular punctures both of same type $\cP_k$. This leads to the $4d$ $\cN=2$ asymptotically free gauge theory $\su(n)+2k\F$ \cite{Nanopoulos:2010zb}. We know from the gauge theory viewpoint that $\cL$ is trivial if $k>0$, from which we can bootstrap that an element $\alpha$ of $\wh{Z}\simeq\Z_n$ wrapped along the cycle $W$ displayed in figure \ref{fig:WH} can be contracted to a trivial loop. In other words, we learn that any element $\alpha$ of $\wh{Z}\simeq\Z_n$ can be moved across an irregular puncture of type $\cP_k$ if $1\le k\le n-1$, see figure \ref{fig:PkMove}. Thus, as far as considerations about 1-form symmetry are concerned, an untwisted irregular puncture of type $\cP_k$ for $k>0$ behaves exactly like an untwisted regular puncture, i.e. such an untwisted irregular puncture can be neglected when determining the 1-form symmetry.

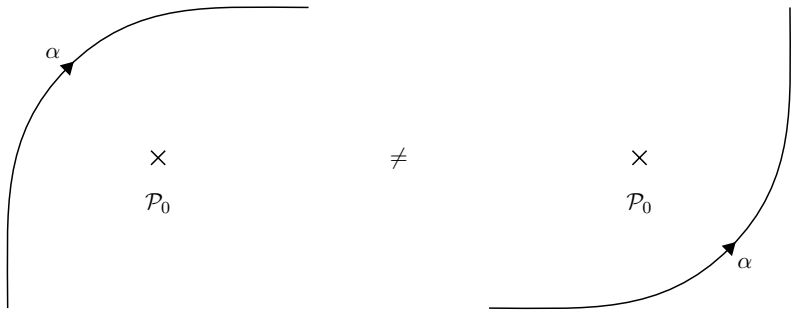
\begin{figure}
\centering
\scalebox{.8}{\begin{tikzpicture}
	\begin{pgfonlayer}{nodelayer}
		\node [style=none] (0) at (-6.5, -1) {};
		\node [style=none] (1) at (-1.5, 4) {};
		\node [style=none] (2) at (1.5, -1) {};
		\node [style=none] (3) at (6.5, 4) {};
		\node [style=NodeCross] (4) at (4, 1.5) {};
		\node [style=NodeCross] (5) at (-4, 1.5) {};
		\node [style=none] (6) at (-5.5, 3) {};
		\node [style=none] (7) at (5.5, 0) {};
		\node [style=UpLefttriangle] (8) at (-5.5, 3) {};
		\node [style=UpLefttriangle] (9) at (5.5, 0) {};
		\node [style=none] (11) at (-4, 0.75) {$\mathcal{P}_{0}$};
		\node [style=none] (12) at (4, 0.75) {$\mathcal{P}_{0}$};
		\node [style=none] (13) at (0, 1.5) {$\neq$};
		\node [style=none] (14) at (-5.75, 3.25) {$\alpha$};
		\node [style=none] (15) at (5.75, -0.25) {$\alpha$};
	\end{pgfonlayer}
	\begin{pgfonlayer}{edgelayer}
		\draw [style=ThickLine, in=-135, out=90] (0.center) to (6.center);
		\draw [style=ThickLine, in=-180, out=45] (6.center) to (1.center);
		\draw [style=ThickLine, in=-135, out=0] (2.center) to (7.center);
		\draw [style=ThickLine, in=-90, out=45] (7.center) to (3.center);
	\end{pgfonlayer}
\end{tikzpicture}
}
\caption{A line carrying $0\neq\alpha\in\wh{Z}$ \emph{cannot} be deformed across an irregular puncture of type $\cP_0$.}
\label{fig:P0Move}
\end{figure}

On the other hand, for $k=0$, we obtain the $4d$ $\cN=2$ pure $\su(n)$ gauge theory. This gauge theory has
\be
\cL=\Z_n\times\Z_n
\ee
with the first $\Z_n$ factor arising from Wilson line operators, and the second $\Z_n$ factor arising from 't Hooft line operators. The $\Z_n$ factor associated to Wilson line operators can be understood as arising from $6d$ surface operators wrapping the cycle $W$ shown in figure \ref{fig:WH}. This identification can be made by observing that when $W$ is very small, we first reduce the $6d$ theory to $5d$ $\cN=2$ $\su(n)$ SYM and then reduce this $5d$ theory to the $4d$ $\cN=2$ pure $\su(n)$ gauge theory due to the presence of boundary conditions associated to the two irregular punctures. In this reduction the $6d$ surface operators wrapping $W$ become line operators in the $5d$ theory and hence can be identified with the Wilson line operators of the $5d$ theory, and then subsequently as Wilson line operators of the $4d$ theory. This means that no element $\alpha$ of $\wh{Z}\simeq\Z_n$ wrapped along the cycle $W$ can be contracted to a trivial loop. Hence, an untwisted irregular puncture of type $\cP_0$ does not allow any element $\alpha$ of $\wh{Z}\simeq\Z_n$ to be moved across it, as  shown in figure \ref{fig:P0Move}.

Now one can ask what is the interpretation of the $\Z_n$ factor associated to 't Hooft line operators from the point of view of the compactification of the $6d$ theory. We propose that this is associated to elements of $\wh{Z}$ inserted along the oriented segment labeled $H$ in figure \ref{fig:WH}. One can then observe that
\be
\langle f(W),f(H)\rangle=\frac1n\,.
\ee
That is, the pairing between elements of $\cL$ obtained by wrapping generator $f$ of $\wh{Z}$ along $W$ and $H$ is precisely the Dirac pairing between fundamental Wilson and 't Hooft operators in the gauge theory.

Notice that our above proposal for 't Hooft line operators implies that a $6d$ surface operator can end on the codimension-2 defect associated to an irregular puncture of type $\cP_0$. This is our first example of a puncture having this property. One would imagine that more general irregular punctures discussed in \cite{Xie:2012hs,Wang:2015mra,Wang:2018gvb} allow a subgroup of $\wh{Z}$ to end on them, depending on the type of puncture. We defer a more thorough analysis to a future work, but finish this section by substantiating our proposal for the properties of punctures of type $\cP_k$ by studying the following example.

\vspace{6mm}

\ni\ubf{Example}: Consider compactifying $A_1$ $(2,0)$ theory on $C_g$ with $n_1$ regular punctures, $n_2$ punctures of type $\cP_1$ and $n_3$ punctures of type $\cP_0$. From our above analysis, we expect
\be\label{AFC}
\cL\simeq\big(\Z_2^{n_3-1}\times\Z_2^g\big)_A\times\big(\Z_2^{n_3-1}\times\Z_2^g\big)_B\,.
\ee
In a particular degeneration limit, we obtain the following $4d$ $\cN=2$ asymptotically free gauge theory
\be
\scalebox{0.95}{
}
\ee
where a trivalent vertex denotes a half-hyper in trifundamental representation, and $n_2,n_3$ count the number of such half-trifundamentals. From the above gauge theory description one can verify that $\cL$ is indeed given by (\ref{AFC}).

\section{1-Form Symmetries from Type IIB Realization}\label{sec:IIB}

\subsection{Class S from Type IIB}

Class S theories can also have a realization in terms of a dual, Type IIB compactification, using geometric field theory methods, developed for general $\mathcal{N}=2$ theories, predating class S \cite{Klemm:1996bj}. 
Type IIB on a canonical singularity gives rise to $\mathcal{N}=2$ SCFTs, and more generally can provide a way to engineer gauge theories. 
The Calabi-Yau $X$ geometries that realize class S theories, can be constructed as ALE-fibrations over a curve
\be
\widetilde{\mathbb{C}^2/\Gamma_{\text{ADE}}} ~\hookrightarrow~ X ~\rightarrow~ C_{g,n}\,,
\ee
where the resolutions parametrized for the ALE-fiber are encoded in a Higgs field $\varphi$. 
The connection is made through the Higgs bundle, \cite{Noguchi:1999xq, Gaiotto:2009hg}. 
The Higgs field $\varphi$ is a meromorphic 1-form valued in the respective ADE Lie algebra $\mathfrak{g}$. 
We consider the 6d (2,0) theory of type ADE on $C_{g,n}$, with the standard topological twist that retains $\mathcal{N}=2$ supersymmetry in 4d, i.e. $SO(5)\rightarrow SU(2) \times U(1)_R$ and $SO(6)\rightarrow SO(4) \times U(1)_L$ twisting the $U(1)_L$ by combining it with the $U(1)_R$ R-symmetry transformation. 
The scalars give rise to the $(1,0)$ and $(0,1)$ forms $\varphi$ and $\bar\varphi$. These define together with the gauge field components (along the curve) the Higgs bundle, satisfying the Hitchin equations. 
The spectral equation defines the SW curve inside the co-tangent bundle of $C_{g,n}$
\be
\text{det} (\varphi -\lambda \,\text{Id}) =0 \,.
\ee
We assume that the Higgs bundle is diagonalizable, i.e. $\varphi = \text{diag} (\lambda_1, \cdots, \lambda_r)$.
The spectral data encodes a local Calabi-Yau, which defines an ALE-fibration over $C$. 
Each sheet is labeled by a fundamental weight of $\mathfrak{g}$. For simplicity let us focus on the $A_{N-1}$ case. There are $N$ sheets, associated to the $L_i$, $i=1, \cdots, N$ fundamental weights, with the simple roots realized as $\alpha_i = L_i - L_{i+1}$. The Higgs field eigenvalues $\lambda_i$ encode the volumes of the rational curve in the ALE-fibration, where each simple root is associated to a rational curve $\mathbb{P}^1_i$, whose volume is determined by 
\be
\int_{\mathbb{P}^1_i}\Omega = \lambda_i- \lambda_{i+1}  \,.
\ee
When $\lambda_i=0$ for all $i$, the full $SU(N)$ symmetry is restored. 
More precisely, the spectral curve allows us to construct three-cycles as follows: if $b_{\alpha}$ are the branch points of the spectral curve, where two sheets of the cover collide, we can construct an $S^3$ by considering the ALE-fiber over the line $\ell_{\alpha, \beta}$ connecting two branch-points in $C$. At each of the branch-points a 2-sphere collapses, and thus we obtain an $S^3$. These three-spheres are Lagrangian and give rise in IIB to the hypermultiplets in 4d. 
Other three-cycles with the topology of $S^2\times S^1$ are obtained by considering the rational curves fibered over closed 1-cycles in the base, which correspond to vectors.

Regular, untwisted punctures correspond to simple poles of $\varphi$. In the ALE-fibration, this maps to sending the volumes of (some) $\mathbb{P}^1$s to infinity. 
The punctures are labeled by partitions of $N = \sum n_ih_i$, where $n_i$ is the multipliticy of the box of height $h_i$ in the Young tableaux. The flavor symmetry is  $G_F=S(\prod_i U(n_i))$. 
E.g. the full punctures corresponding to the partition $1^N$ the flavor symmetry is $SU(N)$, corresponds to sending all $N$ sheets to infinity with the same rate, parameterized by the residue of the pole of $\varphi$.

\begin{figure}
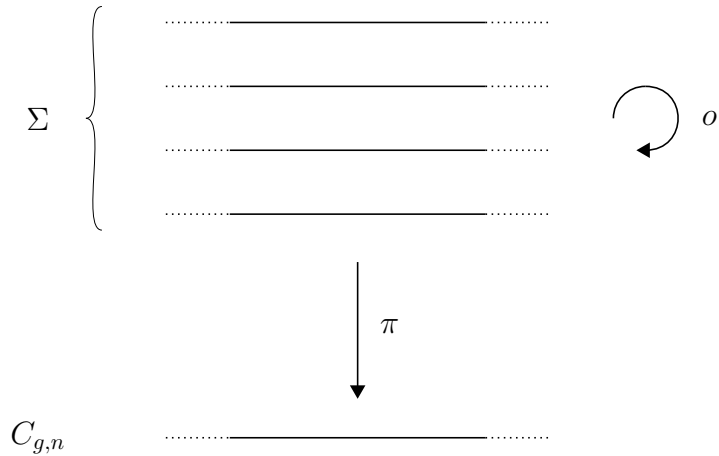

\centering
\scalebox{.85}{
}
\caption{The outer automorphism \eqref{eq:ActionOuterAut} acting on the spectral cover.}
\label{fig:SU4}
\end{figure}

Open and closed twist lines alter the global structure of the ALE geometry. Open twist lines are inserted between punctures and closed twist lines are wrapped along a 1-cycle $B$ of the base $C$, both are labelled by an element $o$ of the outer automorphism group. When encircling a puncture or traversing a 1-cycle intersecting $B$ the Higgs field is acted on the by the outer automorphism $o$, see figure \ref{fig:SU4}. In the ALE-fibration, rational curves $\P^1$ locally sweeping out distinct three-cycles are identified reducing the total number of 3-cycles in $X$. The Poincar\'e dual of these three-cycles are used to expand the supergravity four-form and construct the gauge bosons of the effective 4d theory. The gauge algebra of the theory is therefore determined by the initial choice of ADE gauge group and twist line structure.

\paragraph{Example:} Consider the 6d $(2,0)$ $A_{2n-1}$ theory compactified on the torus $C_g=T^2$ with a closed $b$ twist line along the $B$ cycle. The spectral cover $\Sigma$ is a $2n$-sheeted cover of the torus $T^2$. Each sheet can be thought of as associated to a fundamental weight $L_i$, $i= 1, \cdots, 2n$, and the outer automorphism acts as 
\be\label{eq:ActionOuterAut}
o:\qquad L_{i} \ \longleftrightarrow \  -L_{2n+1-i} \,,
\ee
which induces an action on the simple roots $\alpha_i = L_{i} - L_{i+1} \leftrightarrow \alpha_{2n -i}$. The root $\alpha_n$ is fixed. There are $n$ 3-cycles, one for each orbit of the outer automorphism on the $\P^1$ fibers which determine the root system of the 4d gauge algebra. These 3-cycles intersect linearly with the 3-cycle corresponding to the fixed $\P^1$ lying at the end of this chain.
The root originating from this $\P^1$ is shorter than than the remaining $n-1$ roots and we find the roots system of type $B_n$. Overall we find the gauge group to reduce from $SU(2n)$ to Spin$(2n+1)$ when introducing the twist line, the center of Spin$(2n+1)$ is $\Z_2$.

\subsection{Line operators from IIB}\label{JKP}
The line operators in this context are realized in terms of wrapped D3-branes, on non-compact three-cycles, modulo screening by particles, which are D3-branes wrapped on compact three-cycles. To study these, consider the analog arguments as in \cite{Morrison:2020ool, Closset:2020scj,  Albertini:2020mdx}.
In relative homology, where $\partial X$ is the boundary link 5-fold of the Calabi-Yau three-fold, the line operators are thereby realized in terms of 
\be
\cL = {H_3 (X, \partial X, \mathbb{Z}) \over H_3 (X, \mathbb{Z}) }\,.
\ee
Chasing this through the long exact sequence in relative homology, 
\be
\cdots \ \longrightarrow  \
H_3 (X, \mathbb{Z})\ \stackrel{q}{\longrightarrow} \
H_3 (X, \partial X, \mathbb{Z})\ \stackrel{\partial}{\longrightarrow}  \
H_2 ( \partial X, \mathbb{Z})\ \stackrel{\iota}{\longrightarrow}  \ H_2 (X, \mathbb{Z})\ \stackrel{}{\longrightarrow}  \ \cdots  \,,
\ee
we find that
\be
\cL = {H_3 (X, \partial X, \mathbb{Z}) \over H_3 (X, \mathbb{Z}) }= {H_3 (X, \partial X, \mathbb{Z}) \over \text{Im}(q) } = \text{Im} (\partial) = \text{Ker} (\iota)\,.
\ee
In particular we can write it as 
\be\label{eq:2cycles}
\cL = \{ \ell \in  H_2 (\partial X , \mathbb{Z})|  ~\text{$\ell$ is a 2-cycle in $\partial X$ which becomes trivial in $X$}\} \,.
\ee
The pairing on $\cL$ governing the mutual non-locality of $4d$ line operators descends straightforwardly from the linking pairing on $H_2 (\partial X , \mathbb{Z})$.

The boundary $\partial X$ receives contributions $B_F$ and $B_k$ from the fibers and punctures respectively
\be\label{eq:Boundary}
\partial X_{ C_{g,n}} = B_F \cup \bigcup_{k=1}^n B_k \,,
\ee
where the topology of $B_k$ is given by
\be\label{eq:Bdry}
\widetilde{\mathbb{C}^2/\Gamma_{\text{ADE}}} ~\hookrightarrow~ B_k ~\rightarrow~ S^1\,,
\ee
and the topology of $B_F$ is given by
\be\label{FBC}
S^3/\Gamma_{\text{ADE}} ~\hookrightarrow~ B_F ~\rightarrow~ C_{g,n}\,.
\ee

The contribution of (\ref{FBC}) part of $\partial X_{C_{g,n}}$ to $H_2(\partial X_{ C_{g,n}},\Z)$ is obtained by choosing an element 
$\alpha\in H_1(S^3/\Gamma_{\text{ADE}})$, which is then fibered over a loop $L$ in $C_{g,n}$. We have
\be
H_1(S^3/\Gamma_{\text{ADE}},\Z)\simeq \wh{Z}(\cG)\,,
\ee
where $\cG$ is the simply connected Lie group associated to the ADE Lie algebra $\fg$ associated to $\Gamma_{\text{ADE}}$. Moreover, an outer-automorphism of $\fg$ acts on $H_1(S^3/\Gamma_{\text{ADE}},\Z)$ in precisely the same way as it acts on $\wh{Z}(\cG)$. When the loop $L$ crosses an outer-automorphism twist line $o$, $\alpha$ is transformed to $o\cdot \alpha$. Moreover, any such element $(\alpha,L)\in H_2(B_F,\Z)\subset H_2(\partial X,\Z)$ is clearly trivial, when embedded into $X$ since $\alpha$ is contractible when embedded into $\mathbb{C}^2/\Gamma_{\text{ADE}}$. Thus, contributions of type $(\alpha,L)$ give rise to a non-trivial subgroup 
\be\label{>}
\cL_F\subseteq\cL \,,
\ee
where $\cL$ is defined in (\ref{eq:2cycles}).

Now, notice that the above contributions of the kind $(\alpha,L)$ are precisely the contributions we have been considering throughout the paper. Let us label the group of line operators obtained using the earlier considerations in the paper as $\cL_0$. Then we clearly have
\be\label{<}
\cL_0\subseteq\cL_F \,.
\ee
Thus, the only way for our previous calculation $\cL_0$ and the Type IIB calculation $\cL$ to match is if
\be
\cL_0=\cL_F=\cL \,.
\ee
In the rest of this subsection, we justify this equality.

First thing we need to show is that the contribution of all boundary components $B_k$ to $\cL$ is trivial. Indeed, the only 2-cycles in $B_k$ are the exceptional $\P^1$s in $\widetilde{\mathbb{C}^2/\Gamma_{\text{ADE}}}$, but none of these 2-cycles is trivial when embedded into $X$, and hence do not contribute to $\cL$.

\begin{figure}
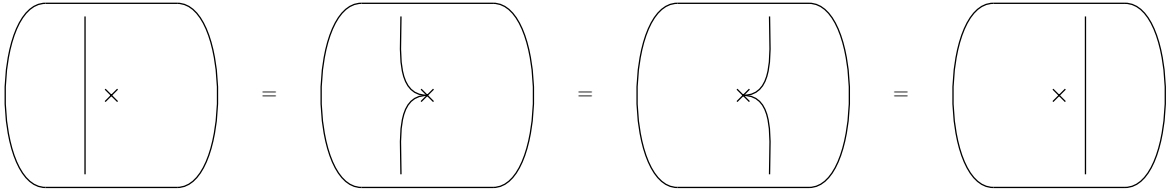

\centering
\scalebox{.70}{
}
\caption{Consider an untwisted regular puncture and a boundary cycle $(L, \alpha)\in H_2(B_F, \mathbb{Z})$, with $\alpha\in H_1(S^3/\Gamma_{\text{ADE}})$. We illustrate how the untwisted puncture does not affect this contribution to the defect group. Left: A line $L$, associated to $(L, \alpha)$. Right: A line $L'$ associated to $(L',\alpha)$. Center-left: Limiting configuration as $L$ is moved towards an untwisted regular puncture. Center-right: Limiting configuration as $L'$ is moved towards the puncture.}
\label{fig:RegularPunctureIrrev}
\end{figure}

Next, we need to show that $(L, \alpha)$ and $(L',\alpha)$ give rise to the same element in $H_2(\partial X,\Z)$ if $L'$ is obtained from $L$ by passing it over an untwisted regular puncture. Consider the limiting configuration of $L$ approaching an untwisted regular puncture $k$, say from the left in figure \ref{fig:RegularPunctureIrrev}. We hit the boundary component $B_k$ at a particular point $p\in S^1$. The fiber component $\alpha$ embeds into the fiber $\big(\widetilde{\mathbb{C}^2/\Gamma_{\text{ADE}}}\big)_p$ of $B_k$ at $p$ via the inclusion map
\be\label{eq:Embedding}
\iota_p\,:\qquad S^3/\Gamma_{\text{ADE}} ~\xhookrightarrow~ \big(\widetilde{\mathbb{C}^2/\Gamma_{\text{ADE}}}\big)_p \,.
\ee
Similarly, the limiting configuration of $L'$ approaching an untwisted regular puncture $k$, say from the right in figure \ref{fig:RegularPunctureIrrev}, hits the boundary component $B_k$ at a particular point $p'\in S^1$. The fiber component $\alpha$ embeds into the fiber $\big(\widetilde{\mathbb{C}^2/\Gamma_{\text{ADE}}}\big)_{p'}$ of $B_k$ at $p'$ via the inclusion map described above. Since the two embeddings of $\alpha$ into $\big(\widetilde{\mathbb{C}^2/\Gamma_{\text{ADE}}}\big)_p$ and $\big(\widetilde{\mathbb{C}^2/\Gamma_{\text{ADE}}}\big)_{p'}$ respectively are homotopic to each other, we deduce that $(L, \alpha)=(L', \alpha)$ as elements of $H_2(\partial X,\Z)$. 

\begin{figure}
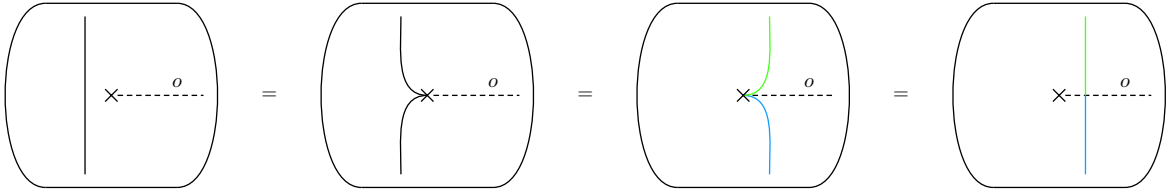

\centering
\scalebox{.70}{
}
\caption{Consider again $(L, \alpha), (L',\alpha) \in H_2(B_F, \mathbb{Z})$ with $\alpha\in H_1(S^3/\Gamma_{\text{ADE}})$.  Left: A line $L$ associated to $(L, \alpha)$. Right: A line $L'$ associated to $(L', \alpha)$ along the blue subsegment and $o\cdot\alpha$ along the green subsegment. Center-left: Limiting configuration as $L$ is moved towards an untwisted regular puncture. Center-right: Limiting configuration as $L'$ is moved towards the puncture. The central equality only holds for $\alpha=o\cdot\alpha$.}
\label{fig:TwistedPunctureRev}
\end{figure}

Finally, we need to show that $(L, \alpha)$ and $(L', \alpha)$ give rise to the same element in $H_2(\partial X,\Z)$ if $L'$ is obtained from $L$ by passing it over an twisted regular puncture, as long as $\alpha$ is left invariant by the action of the outer-automorphism associated to the twist line emanating from the twisted regular puncture. The argument proceeds exactly as in the untwisted case since the twist line is immaterial if $\alpha$ is left invariant by the corresponding outer-automorphism action. On the other hand, if $\alpha$ is not left invariant by the outer-automorphism, then $L'$ needs to be divided into two sub-rays (denoted by blue and green respectively in figure \ref{fig:TwistedPunctureRev}) with $\alpha$ inserted along the blue sub-ray and $o\cdot\alpha$ inserted along the green sub-ray. In particular, there is no consistent limiting configuration as $L'$ approaches the puncture, and the above argument fails. Thus, $L$ and $L'$ give rise to different elements of $H_2(\partial X,\Z)$ (and hence $\cL$) if $\alpha$ is acted upon by the twist line emanating from the regular puncture.

The above argument can be viewed as a justification of our key assumption used in the earlier parts of the paper: If $L$ is a loop surrounding a regular (untwisted or twisted) puncture carrying an element $\alpha\in\wh{Z}(\cG)$ left invariant by the twist line emanating from the puncture, then such a loop is trivial in $\cL$.
As an alternative approach one might consider arguing that closing an untwisted regular puncture does not change the defect group. It would be interesting to develop this point of view. Here we note, that in the geometric descriptions one could argue as folllows: regular punctures characterize base points at which fibral $\P^1$'s both decompactify and potentially braid upon. For line operators the decompactification of cycles is immaterial. We can therefore rescale Higgs field with a factor of the base coordinate $z$ preserving the braiding structure. This completely removes regular punctures. In other words, regular punctures can be filled in from the perspective of line operators and do not contribute to the defect group. It would be interesting to develop the precise dictionary, and to expand it to include irregular punctures.

Generically the above procedure can be applied to any canonical singularity. E.g. even in the case of general irregular punctures, which realize 
Argyres Douglas theories, that do not necessarily admit a Lagrangian description. The theories of type $\text{AD}[G, G']$ have a realization in terms of Type IIB on a canonical singularity and for $\text{AD}[G, G']$ theories, the 1-form symmetries are non-trivial only for $G=A_N$ with $N>1$ and $G' = D, E$ type, see  \cite{Closset:2020scj, Albertini:2020mdx}. These results should provide further insights into computing the one-form symmetry for irregular punctures more generally. 


\section*{Acknowledgements}
We thank Fabio Apruzzi, Chris Beem, Cyril Closset, Po-Shen Hsin, Du Pei, Yuji Tachikawa, Yi-Nan Wang and Gabi Zafrir for discussions. We in particular thank Yuji Tachikawa for detailed comments on the draft.\\
This work is supported by ERC grants 682608 (LB and SSN) and 787185 (LB). The work of MH is supported by the Studienstiftung des Deutschen Volkes. 
SSN also acknowledges support through the Simons Foundation Collaboration on "Special Holonomy in Geometry, Analysis, and Physics", Award ID: 724073, Schafer-Nameki.

\appendix 

\section{Summary of Notation}
\label{app:Not}
\bit
\item$\fg$: Mostly denotes the A,D,E Lie algebra denoting the $6d$ $\cN=(2,0)$ theory under consideration. Can also denote a simple gauge algebra (simply or non-simply laced) for a $4d$ $\cN=2$ gauge theory depending on the context.
\item$\cG$: The simply connected group associated to a simple Lie algebra $\fg$.
\item$Z(\cG)$: The center of a simply connected group $\cG$.
\item$\wh{Z}(\cG)$: The Pontryagin dual of the center of a simply connected group $\cG$. For a $6d$ $\cN=(2,0)$ theory associated to an A,D,E Lie algebra $\fg$, $\wh{Z}(\cG)$ captures the group of dimension-2 surface operators modulo screenings, also known as the \emph{defect group} of the $6d$ theory.
\item$C_g$: A Riemann surface of genus $g$ which might carry punctures depending on the context.
\item$\cL$: The set of line operators (modulo screenings and flavor charges) for a \emph{relative} $4d$ $\cN=2$ theory obtained by compactifying a $6d$ $\cN=(2,0)$ theory on a Riemann surface $C_g$, possibly in the presence of twist lines and (untwisted and twisted) regular punctures.
\item$\langle\cdot,\cdot\rangle$: Often referred to as \emph{pairing}. It takes two elements of $\wh{Z}(\cG)$ or of $\cL$, and outputs an element of $\R/\Z$ which captures the phase associated to mutual non-locality of the defect operators associated to the two elements.
\item$\Lambda$: Often referred to as \emph{polarization} or \emph{maximal isotropic subgroup}. This is a maximal subgroup of $\cL$ such that the pairing $\langle\cdot,\cdot\rangle$ on $\cL$ restricted to this subgroup $\Lambda$ vanishes. A choice of such a $\Lambda$ is correlated to the choice of an \emph{absolute} $4d$ $\cN=2$ theory.
\item$\wh{\Lambda}$: Pontryagin dual of polarization $\Lambda$. Captures the 1-form symmetry of the absolute $4d$ $\cN=2$ theory associated to a polarization $\Lambda$.
\item$\F$: Denotes fundamental representation for $\fg=\su(n),\sp(n)$; vector representation for $\fg=\so(n)$; representations of dimension $\mathbf{7},\mathbf{26},\mathbf{27},\mathbf{56}$ for $\fg=\fg_2,\ff_4,\fe_6,\fe_7$ respectively; and the adjoint representation for $\fg=\fe_8$.
\item$\mathsf{\Lambda}^n$: Denotes the $n$-index antisymmetric \emph{irreducible} representation for $\fg=\su(n),\sp(n)$.
\item$\S^2$: Denotes the $2$-index symmetric representation for $\fg=\su(n)$.
\item$\S$: Denotes the irreducible spinor representation for $\fg=\so(n)$.
\item$\C$: Denotes the irreducible cospinor representation for $\fg=\so(2n)$.
\item$n\mathsf{R}$: Denotes $n$ full hypermultiplets transforming in a representation $\mathsf{R}$.
\item$\frac{2n+1}2\mathsf{R}$: Denotes $n$ full hypermultiplets and a half-hypermultiplet transforming in a pseudo-real representation $\mathsf{R}$.
\eit

\bibliographystyle{JHEP}
\bibliography{ref}

\end{document}